\documentclass[12pt,preprint]{aastex}
\newcommand{\dmm}{\mbox{$\Delta$$m$$_{15}(B)$}}

\def\mbar{\ifmmode\overline{m}\else$\overline{m}$\fi}
\def\Mbar{\ifmmode\overline{M}\else$\overline{M}$\fi}
\def\mibar{\ifmmode\overline{m}_I\else$\overline{m}_I$\fi}
\def\MIbar{\ifmmode\overline{M}_I\else$\overline{M}_I$\fi}
\def\Nbar{\ifmmode\overline{N}\else$\overline{N}$\fi}

\def\ho{\ifmmode H_0\else$H_0$\fi}

\def\dmod{\ifmmode(m{-}M)_0\else$(m{-}M)_0$\fi}
\def\mM{\ifmmode(m{-}M)_0\else$(m{-}M)_0$\fi}
\def\vi{\ifmmode(V{-}I)\else$(V{-}I)$\fi}
\def\viz{\ifmmode(V{-}I)_0\else$(V{-}I)_0$\fi}
\def\EBV{\ifmmode E_{B-V}\else$E_{B-V}$\fi}

\shorttitle{Optical Spectroscopy of Type Ia Supernovae}
\shortauthors{Matheson et al.}

\begin{document}
\received{} 
\title{Optical Spectroscopy of Type Ia Supernovae\altaffilmark{1}}

\author{T. Matheson,\altaffilmark{2}
R.~P. Kirshner,\altaffilmark{3}
P. Challis,\altaffilmark{3}
S. Jha,\altaffilmark{4}
P.~M. Garnavich,\altaffilmark{5}
P. Berlind,\altaffilmark{6}
M.~L. Calkins,\altaffilmark{6}
S. Blondin,\altaffilmark{3}
Z. Balog,\altaffilmark{7}
A.~E. Bragg,\altaffilmark{8}
N. Caldwell,\altaffilmark{3}
K. Dendy~Concannon,\altaffilmark{9}
E.~E. Falco,\altaffilmark{6}
G.~J.~M. Graves,\altaffilmark{10}
J.~P. Huchra,\altaffilmark{3}
J. Kuraszkiewicz,\altaffilmark{3}
J.~A. Mader,\altaffilmark{11}
A. Mahdavi,\altaffilmark{12}
M. Phelps,\altaffilmark{3}
K. Rines,\altaffilmark{3}
I. Song,\altaffilmark{13}
and B.~J. Wilkes\altaffilmark{3}
}

\nopagebreak 

\altaffiltext{1}{\vspace{0.00cm}Based in part on observations obtained
  at the F.~L. Whipple Observatory, which is operated by the
  Smithsonian Astrophysical Observatory, and the MMT Observatory, a joint
  facility of the Smithsonian Institution and the University of
  Arizona.}

\altaffiltext{2}{\vspace{0.00cm}National Optical Astronomy
  Observatory, 950 N. Cherry Avenue, Tucson, AZ 85719-4933;
  {matheson@noao.edu}}

\altaffiltext{3}{\vspace{0.00cm}Harvard-Smithsonian Center for
  Astrophysics, 60 Garden Street, Cambridge, MA 02138;
  {kirshner@cfa.harvard.edu}, {pchallis@cfa.harvard.edu},
  {sblondin@cfa.harvard.edu},
  {ncaldwell@cfa.harvard.edu}, 
  {jhuchra@cfa.harvard.edu}, {jkuraszkiewicz@cfa.harvard.edu},
  {mphelps@cfa.harvard.edu}, {krines@cfa.harvard.edu}}

\altaffiltext{4}{\vspace{0.00cm}Kavli Institute for Particle
  Astrophysics and Cosmology, P.O. Box 20450, Stanford, CA  94309;
  {saurabh@slac.stanford.edu}}

\altaffiltext{5}{\vspace{0.00cm}Department of Physics, 
    University of Notre Dame, 
    225 Nieuwland Science Hall, Notre Dame, IN 46556-5670;
  {pgarnavi@nd.edu}}

\altaffiltext{6}{\vspace{0.0cm}F.~L.~Whipple Observatory, 670
Mt.~Hopkins Road, P.O.~Box 97, Amado, AZ 85645;
{pberlind@cfa.harvard.edu}, {mcalkins@cfa.harvard.edu},
{efalco@cfa.harvard.edu} }

\altaffiltext{7}{\vspace{0.0cm}Steward Observatory, University of
Arizona, 933. N. Cherry Avenue Tucson AZ 85721 (on leave from the
Department of Optics and Quantum Electronics, University of Szeged,
D{\'o}m~t{\'e}r 9, H-6720 Szeged, Hungary); {zbalog@as.arizona.edu}}

\altaffiltext{8}{\vspace{0.0cm} Department of Physics and Astronomy,
  Bowling Green State University, Bowling Green, OH
  43403;{aebragg@bgnet.bgsu.edu}}

\altaffiltext{9}{\vspace{0.0cm}Department of Chemistry and Physics,
  King's College, 133 North River Street, Wilkes-Barre, PA 18711;
  {kdconcan@kings.edu}}

\altaffiltext{10}{\vspace{0.0cm}UCO/Lick Observatory, University of
California, Santa Cruz, CA 95064; {graves@ucolick.org}}

\altaffiltext{11}{\vspace{0.0cm}W.~M. Keck Observatory, 65-1120
  Mamalahoa Highway, Kamuela, HI 96743; {jmader@keck.hawaii.edu}}

\altaffiltext{12}{\vspace{0.0cm}Department of Physics and Astronomy,
  University of Victoria, Victoria, BC V8W 3P6, Canada;
  {amahdavi@uvic.ca}}

\altaffiltext{13}{\vspace{0.0cm}Spitzer Science Center, IPAC/Caltech,
  Pasadena, CA 91125; {song@ipac.caltech.edu}}

\begin{abstract}

We present 432 low-dispersion optical spectra of 32 Type Ia supernovae
(SNe~Ia) that also have well-calibrated light curves.  The coverage
ranges from 6 epochs to 36 epochs of spectroscopy.  Most of the data
were obtained with the 1.5m Tillinghast telescope at the F.~L.  Whipple
Observatory with typical wavelength coverage of 3700-7400\AA\ and a
resolution of $\sim7$\AA.  The earliest spectra are thirteen days
before $B$-band maximum; two-thirds of the SNe were observed before
maximum brightness.  Coverage for some SNe continues almost to the
nebular phase.  The consistency of the method of observation and
the technique of reduction makes this an ideal data set for studying
the spectroscopic diversity of SNe~Ia.

\end{abstract}


\setcounter{footnote}{13}

\section{Introduction}

Type Ia supernovae (SNe~Ia) have long been intriguing objects for
astronomers.  As individual objects, they present complex problems
about the nature of their progenitors \citep[e.g., ][ and references
  therein]{howell01a, branch01, nomoto03, stritzinger06}, the physics
of the explosion mechanism\citep[e.g., ][ and references
  therein]{woosley86, hillebrandt00, gamezo04}, and the factors that
produce the observed range of diversity \citep[e.g., ][]{hatano00,
  li01b, benetti05}.  In recent years, a great deal of attention has
been focused on the fact that absolute magnitudes of SNe~Ia can be
deduced from the shape of their light curves \citep[e.g.,
][]{phillips93, hamuy96a, riess96, perlmutter97, jha07}.  Once this
calibration has been applied, SNe~Ia are the best extragalactic
distance measuring tools.  Combined with their large intrinsic
brightness, this makes SNe~Ia extremely valuable as cosmological
distance indicators.  Using SNe~Ia as cosmological lighthouses led to
a recent revolution in cosmology, with the discovery that the Universe
was accelerating \citep[e.g., ][]{riess98, perlmutter99, riess01,
  knop03, tonry03,riess04, astier06, riess07, woodvasey07}, contrary to all expectations.  The nature
of the dark energy that is producing the acceleration is one of the
great unanswered questions of current physics.

For questions associated with understanding individual SNe~Ia, as well
as those related to their use as high-redshift distance indicators,
the quality of the answers will be based upon the underlying data that
are used to make inferences.  There have been much data published about
SNe~Ia, but most have been from detailed studies of individual
objects, with early pioneering work on SN~1972E \citep{kirshner73,
  kirshner75} up to more recent studies  \citep[e.g., ][]{kirshner93,
  stritzinger02, krisciunas03, benetti04, pignata04, kotak05}.  The
extreme examples of SNe~Ia, such as the overluminous SN~1991T
\citep{filippenko92a, phillips92} and the underluminous SN~1991bg
\citep{filippenko92b, leibundgut93}, have also been well studied.  The
drawbacks of earlier samples of SN spectra were the heterogeneous
nature of the data and the relatively small size of the sample.
Examples of the data can be seen at the University of Oklahoma's
supernova spectra database (SUSPECT,
http://bruford.nhn.ou.edu/$\sim$suspect/index1.html) or the web site of the
European Research Training Network on the Physics of Type Ia Supernova
Explosions (http://www.mpa-garching.mpg.de/$\sim$rtn/).  The spectra were
often obtained at a variety of sites and reduced in different ways.
For photometry alone, there are several large, homogeneous data sets
that have been published \citep{hamuy96b, riess99, jha06}, consisting
of light curves of a wide variety of SNe~Ia.  There have been no
large, homogeneous data sets of spectra of SNe~Ia.  Such a sample will
have a wide variety of applications, from testing explosion models to
understanding systematic errors that plague the use of SNe~Ia as
cosmological distance indicators.  Although most attention is focused
on the photospheric-phase spectra, the nebular-phase spectra can
reveal much about SNe~Ia.  Nebular lines in time-series spectra of
SNe~Ia show direct evidence for the changing ratio of cobalt and iron
lines, implying that they are powered by radioactive iron-peak
elements \citep{kuchner94}.  In addition, \citet{mazzali98} showed
that the width of nebular lines was related to the luminosity of the
SN.

Two of the large light-curve data sets mentioned above \citep{riess99,
jha06} are the result of a program begun in 1993 by the SN group at
the Harvard-Smithsonian Center for Astrophysics (CfA) to monitor SNe
(of all types) photometrically and spectroscopically with the
telescopes at the F.~L. Whipple Observatory (FLWO) on Mt. Hopkins,
Arizona.  Through the use of queue scheduling for spectroscopic
observations and a cooperative strategy of a small allocation of
photometric time per night, we have been able to obtain data with a
frequent enough cadence to acquire good data sets on many objects.
The early spectroscopic coverage was mainly for classification, but,
starting in 1997, we began to follow objects in earnest.
Classification is still a major part of the program; between 2000 Sep
and 2003 Sep, we classified 39\% of the low-redshift SNe accessible
from the Northern hemisphere.

\begin{figure}
\epsscale{0.75}
\rotatebox{90}{
\plotone{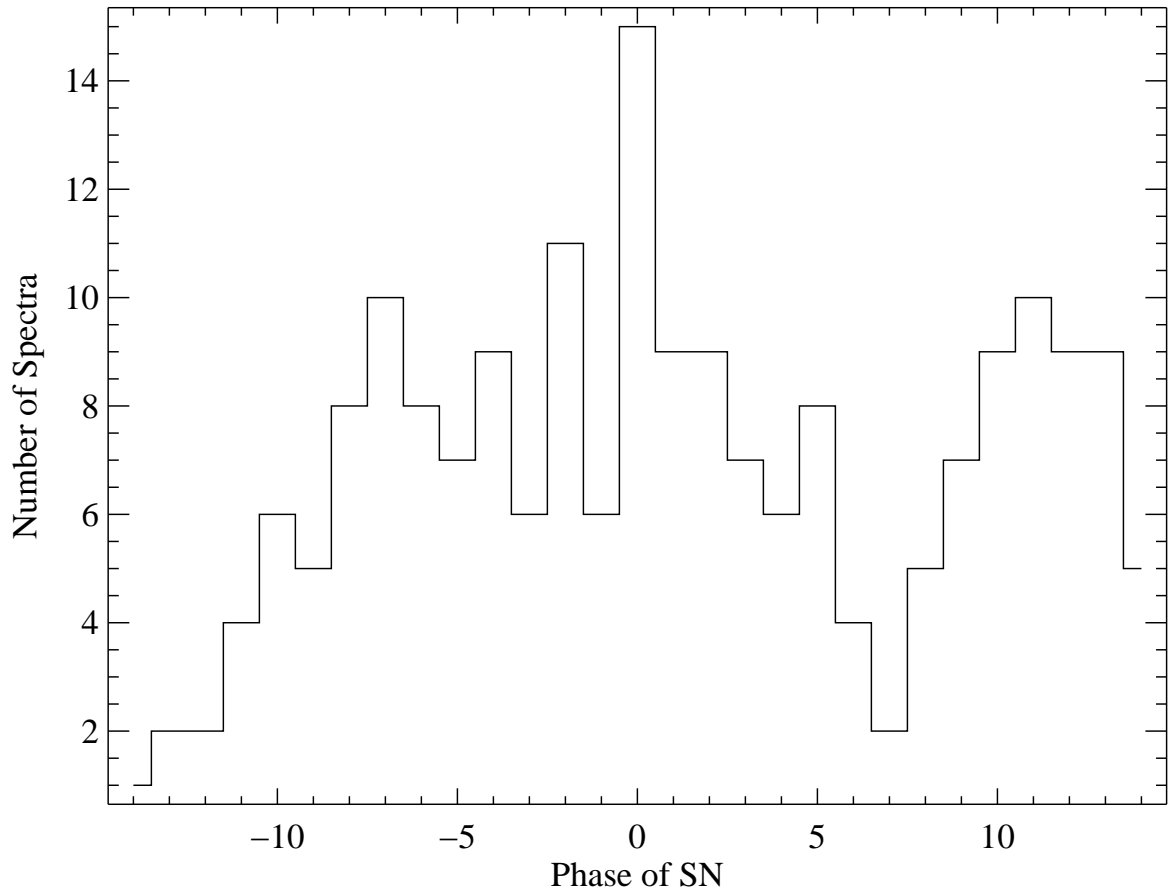}}
\caption{Histogram of the number of individual SN spectra at each
  epoch within fourteen days of maximum brightness.\label{hist}}
\end{figure}

\begin{figure}
\epsscale{0.75}
\rotatebox{90}{
\plotone{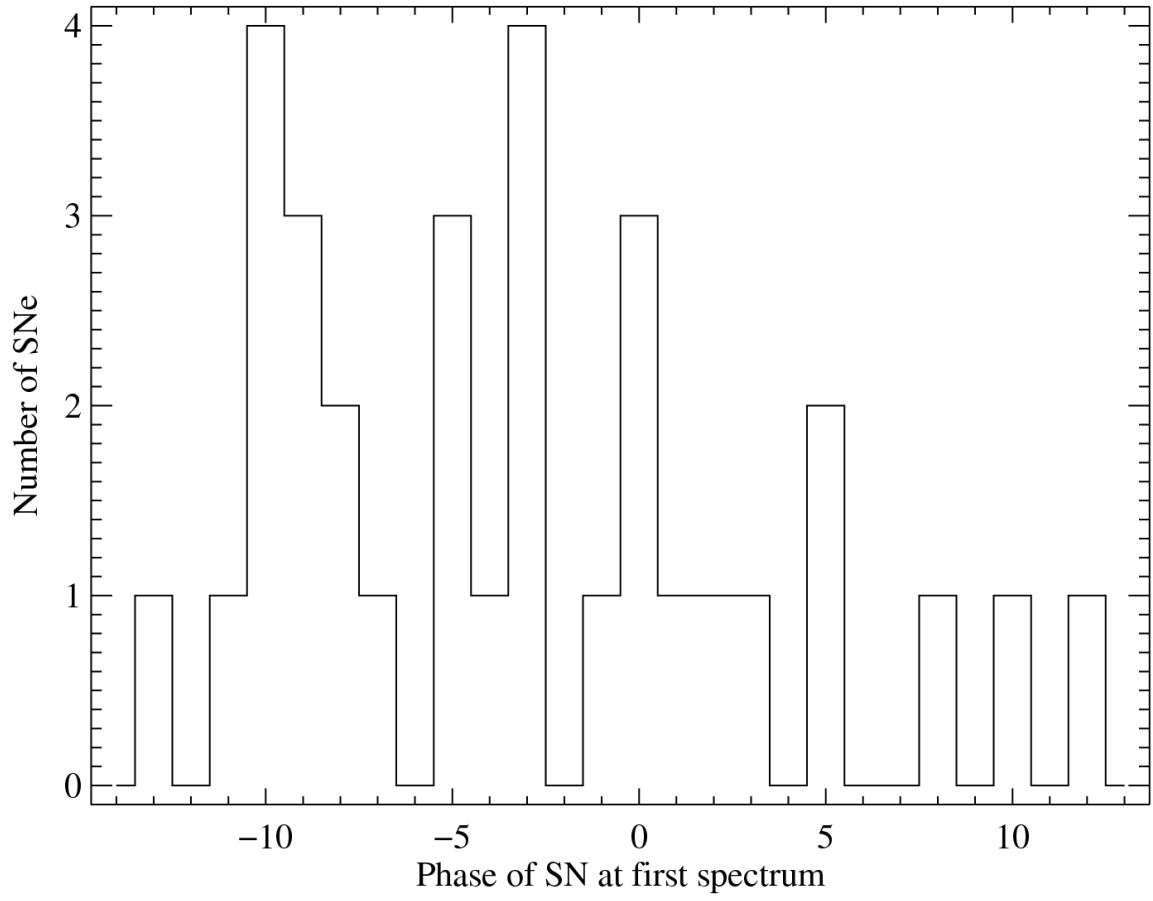}}
\caption{Histogram of the number of SNe at the epoch of the first
  spectrum.\label{first}}
\end{figure}

\begin{figure}
\epsscale{0.75}
\rotatebox{90}{
\plotone{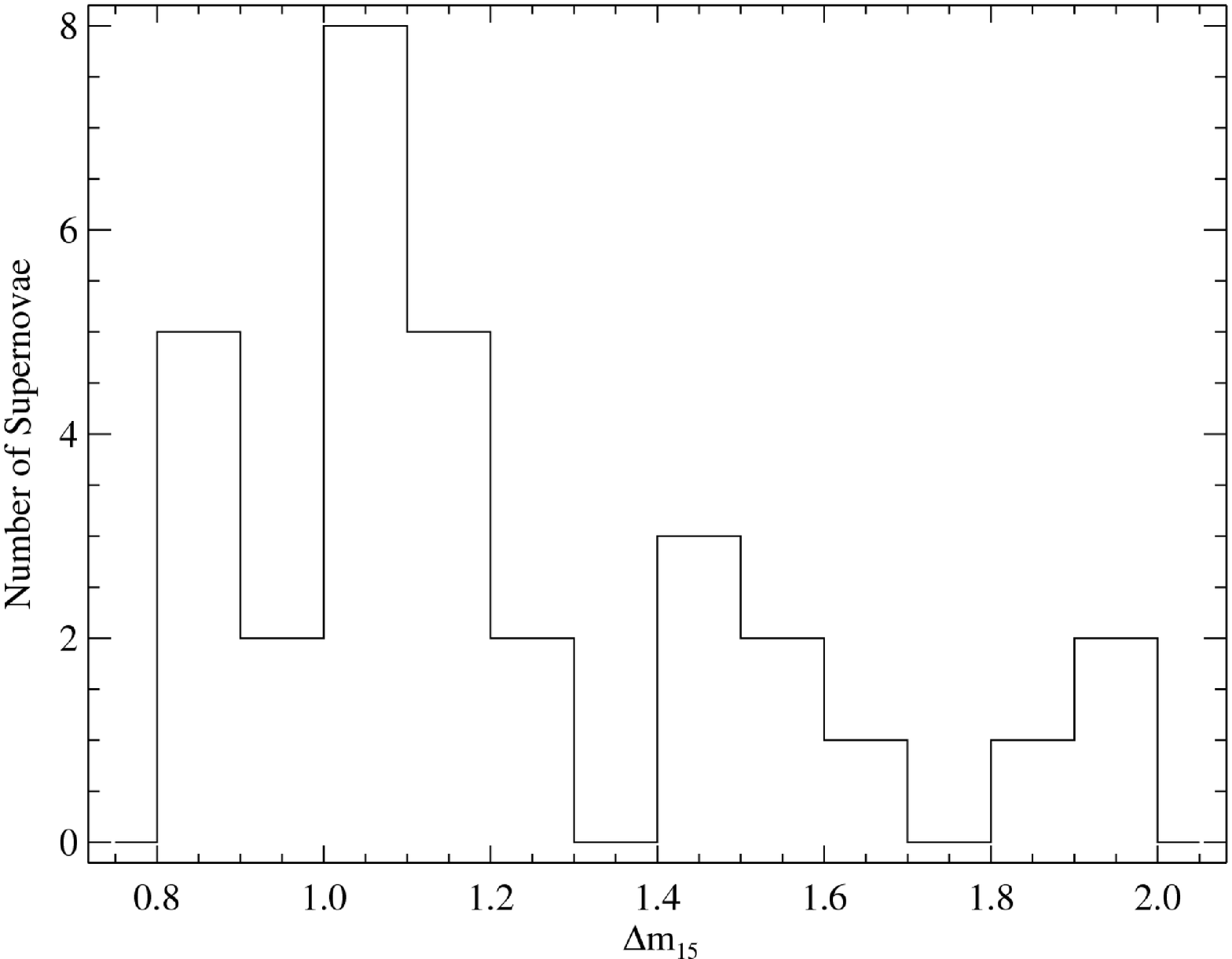}}
\caption{Histogram of the number of SNe versus \dmm\ \citep[From][]
{jha06}.\label{dm15dist}}
\end{figure}

In this paper, we present the first release of some of the
spectroscopic data obtained at Mt. Hopkins by the CfA SN group between
1997 and early 2001.  The decision to follow a specific SN~Ia with
extensive spectroscopic coverage was based upon apparent brightness,
availability of telescope time, and the relative phase of the SN at
our first spectrum.  For the purposes of the sample presented here, we
only include SNe~Ia for which we have a reasonable number of spectra
($>6$).  The final criterion for selecting objects from the CfA SN
database was whether or not there was a calibrated light curve.  With
a light curve, the epoch of maximum is established.  This also gives
us the potential to correlate the light-curve shape with spectroscopic
properties, which is an important goal for this program.  Most of the
photometry for the spectroscopic sample in this paper is presented by
\citet{jha06}.  A few of these SNe~Ia were published as single
objects: \citep[SN~1998aq,][]{riess05}, \citep[SN~1998bu,][]{jha99b},
\citep[SN~1999by,][]{garnavich04}.  Some SNe from 2000 await final
photometric calibration, and so are not included here.  In total,
there were 32 SNe~Ia that fit all the criteria, with 432 individual
spectra.  A histogram of the number of epochs of spectroscopy for the
fourteen days before and after maximum brightness is shown in Figure
\ref{hist}.  Many of the objects were observed well before maximum
brightness.  Figure \ref{first} shows the histogram of the epoch of
the SN at the first spectrum.  In addition, the SNe span the known
range of \dmm\ (Figure \ref{dm15dist}).  The SNe selected, along with
some properties of the host galaxies, are listed in Table
\ref{sndata}.  Virtually all the spectra are from the same instrument,
and they have all reduced in the same manner.  In a companion paper
(Matheson et al. 2008, in preparation), we will present a preliminary
analysis of the spectroscopic characteristics of this data set.  In
that paper, we will relate spectroscopic properties to the light-curve
shapes of the SNe, specifically how the strength of the silicon
features correlates with decline rate.  In addition, we will show the
degree of variation within the SNe~Ia and how this changes with
decline rate.

All the spectra presented in this paper will be made publicly
available through the CfA Supernova Archive
(http://www.cfa.harvard.edu/supernova/SNarchive.html).  This archive
contains all published data from the CfA SN group, both photometric
and spectroscopic.

\clearpage

\clearpage
\section{Observations}

The majority of the spectra presented in this paper were obtained with
the 1.5~m Tillinghast telescope at FLWO using the FAST spectrograph
\citep{fabricant98}.  Spectroscopic observations with the Tillinghast
(both low-dispersion with FAST such as are discussed here and
high-dispersion echelle spectra) are made in a queue-scheduled mode.
Two professional observers (P. Berlind \& M.~L. Calkins) were the primary
observers during the period when these spectra were taken.  In
addition, portions of the telescope schedule are staffed by other CfA
personnel.  The CfA SN group would request two or three observations
per night when FAST was scheduled, subject to constraints at the
telescope (e.g., weather, instrument problems, conflicts with other
programs).  Observational details of the spectra are listed in Table
\ref{obslog}. 

The FAST spectrograph uses a 2688$\times$512 Loral CCD
with a spatial scale of 1.$\arcsec$1 per pixel in the binning mode
used for these observations.  The grating used yields a resolution of
$\sim 7$\AA. The usual setup for observations covered the usable wavelength
range of $\sim$3700 to $\sim$7500~\AA.  Slight variations in the
wavelength range were sometimes introduced during instrument
changes.  Other programs in the FAST queue might occasionally require
different spectrograph settings, resulting in different wavelength
ranges.  In addition, for bright and unusual SNe, we would request
multiple observations with different grating tilts to observe a
broader wavelength range.  The typical slit width was 3$\arcsec$.  For
most of the observations obtained in 1997 and 1998, the slit was
oriented with a position angle of 90$^{\circ}$.  Starting in late 1998,
the slit was typically repositioned to the parallactic angle unless
the object was at a small airmass.

Some spectra were obtained at the 6.5-m MMT Observatory with the
Blue Channel spectrograph \citep{schmidt89}.  These observations
were made during classically scheduled nights, not through
queue-scheduled or interrupt time.  The Blue Channel uses a
2688$\times$512 UA/ITL CCD with a spatial scale of $0.\arcsec6$ per
pixel in the binning mode used for these observations.  The grating
used yields a resolution of $\sim 8$\AA.

\section{Data Reduction}

The FAST data were all reduced in the same consistent manner.  Using
IRAF\footnote{IRAF is distributed by the National Optical Astronomy
Observatory, which is operated by the Association of Universities for
Research in Astronomy, Inc., under cooperative agreement with the
National Science Foundation.}, we would correct for overscan on the
CCD frames and trim the extraneous portions.  In general, the FAST CCD
does not show a bias pattern in zero-time readouts, so we did not
subtract bias frames to avoid introducing additional noise.  In
addition, dark current is not generally a problem with FAST.  There
are a few rare times after UV flashing that the chip has a
dark-current problem, but it is bad enough that it cannot be corrected
and so the affected portion of the spectrum has been trimmed off in
the figures presented herein.  The flat-field frames are combined and
normalized with a low-order spline fit.  The data are then flattened
with these normalized flats.  The spectra were optimally extracted
using the prescription of \citet{horne86} as implemented in the IRAF
\emph{apall} package.  Wavelength calibration was accomplished with
HeNeAr lamps taken immediately after each SN exposure.  A low-order
polynomial was fit to the lines in the calibration lamps, and the
solution applied to the extracted objects.  At a later stage in the
reduction process, we applied small-scale adjustments derived from
night-sky lines in the SN frames.

Once the data were extracted as one-dimensional, wavelength-calibrated
spectra, we used our own routines in IDL to flux calibrate them.  This
entailed spline fits to the standard stars to assign fluxes.  The
relative spectrophotometry is good (see discussion below), but no
attempt was made to put the spectra on an absolute scale.
Spectrophotometric standard stars used for each spectrum are listed in
Table \ref{obslog}.  Using the well-exposed continua of the
spectrophotometric standard stars as smooth-spectrum sources, we
removed telluric features from the spectra using techniques described
by \citet{wade88, bessell99, matheson00a}.

The spectra presented here that were obtained in 1997-1999 were, in
general, not observed at the parallactic angle \citep{filippenko82}.
Most spectra were taken at a small airmass, but some could be affected
by atmospheric dispersion.  Table \ref{obslog} lists the observed
position angle as well as the proper parallactic angle for each
spectrum along with the airmass.  Data taken for the CfA SN group with
FAST from 2000 on were observed at the parallactic angle (unless at
small airmass).

During the period that the observations described here were made, the
optics of the FAST spectrograph blocked blue light, meaning that the
FAST spectra do not suffer from second-order light
contamination\footnote{When we moved the grating tilt to observe at
  red wavelengths, we did use order-blocking filters.}.  The spectra
obtained with the Blue Channel spectrograph can suffer from
second-order contamination, but, through careful cross-calibration
with standard stars of different colors, we have minimized the
problems this might cause.  On any given night, we would try to
observe a relatively blue standard star (typically an sdO) and a
relatively red standard star (typically an sdF).  The sdO standards
provide a better calibration in the blue portion of our spectrum
(below $\sim$4500 \AA) where they will generally have more counts, but
also lack the Balmer jump that can adversely affect calibration.  The
red standard stars will have little blue flux, and thus little
second-order contamination.  Most of our targets, even at relatively
early phases when the spectra can be blue, have little second-order
contamination as well.  Each spectrum is calibrated with both the blue
and the red standard stars.  The blue and red portions of each
spectrum are then joined, typically near 4500 \AA, so that we get a
good calibration of the blue half, without suffering second-order
problems in the red half.  Some residual contamination remains, but
tests with standard stars indicate that we have mitigated most of the
problems.  

\begin{figure}
%\epsscale{0.75}
\plotone{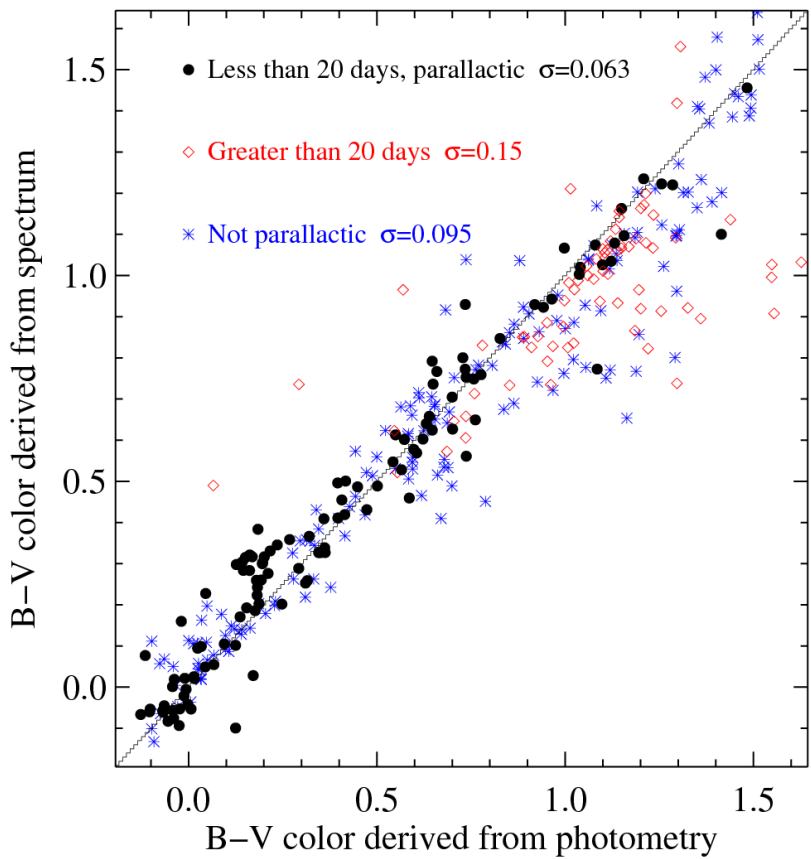}
\caption{Comparison of the colors of the SNe~Ia in our sample derived
  from photometry and spectroscopy.  For objects observed at the
  parallactic angle (or small airmass) and within twenty days of
  $B$-band maximum (\emph{filled, black circles}), the scatter around
  zero difference is 0.063.  For objects observed over twenty days
  after maximum (\emph{open, red diamonds}), the scatter is 0.15.  For
  the early objects in our sample, when we did not consistently
  observe at the parallactic angle (\emph{blue asterisks}), the
  scatter is 0.095.\label{colors}}
\end{figure}

Because the spectra were selected from a sample for which calibrated
light curves exist, we are able to check the accuracy of the relative
spectrophotometry.  We used the light curves of \citet{jha06} to
determine the $B$ and $V$ magnitudes of the SNe at the time of each
spectrum.  When the photometry was not coincident, we interpolated
from the nearest data points.  We then took each spectrum and
convolved it with $B$ and $V$ filter functions in order to derive a
$B-V$ color.  Figure \ref{colors} shows the comparison between the
$B-V$ color based on photometry and the $B-V$ color derived from the
spectra.  For the objects observed from 1997 to late 1999 when we did
not consistently use the parallactic angle, the scatter around zero
difference in color is $\sigma = 0.095$.  The spectra observed at
phases later than twenty days past $B$-band maximum also have a
relatively high scatter of $\sigma = 0.15$.  There are a number of
factors that could cause this difference.  One is that the spectrum
becomes increasingly dominated by line emission as it ages, so that
convolution with a filter that is not precisely matched to the
photometry might introduce a systematic error.  Another important
difference with the later spectra is that they are fainter, so that
host galaxy contamination becomes more significant.  When the spectra
were observed at the parallactic angle during phases when the spectrum
was more continuum-dominated, the scatter is only $\sigma = 0.063$, a
relatively small error.  We believe this indicates how well-calibrated
the spectra in this sample are.
\clearpage
\clearpage

\begin{figure}
\epsscale{0.75}
\rotatebox{90}{
\plotone{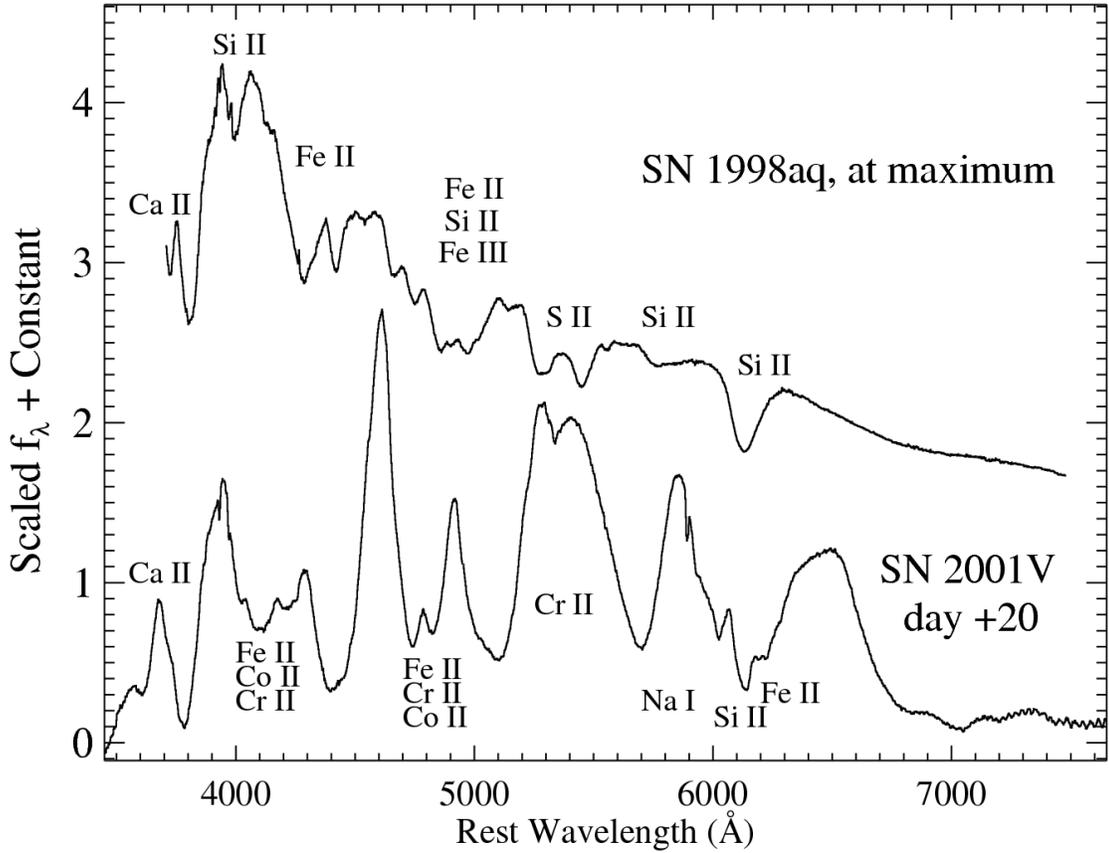}}
\caption{Spectrum of SN~1998aq at $B$-band maximum and spectrum of
   SN~2001V at 20 days past $B$-band maximum.  The flux units are
   f$_{\lambda}$ (ergs s$^{-1}$ cm$^{-2}$ \AA$^{-1}$) that have been
   normalized and then additive offsets applied for clarity. The
   systemic heliocentric velocity listed in Table \ref{sndata} has
   been removed.  Major features of the spectra are
   labeled.\label{labels}}
\end{figure}

\begin{figure}
\epsscale{0.75}
\rotatebox{90}{
\plotone{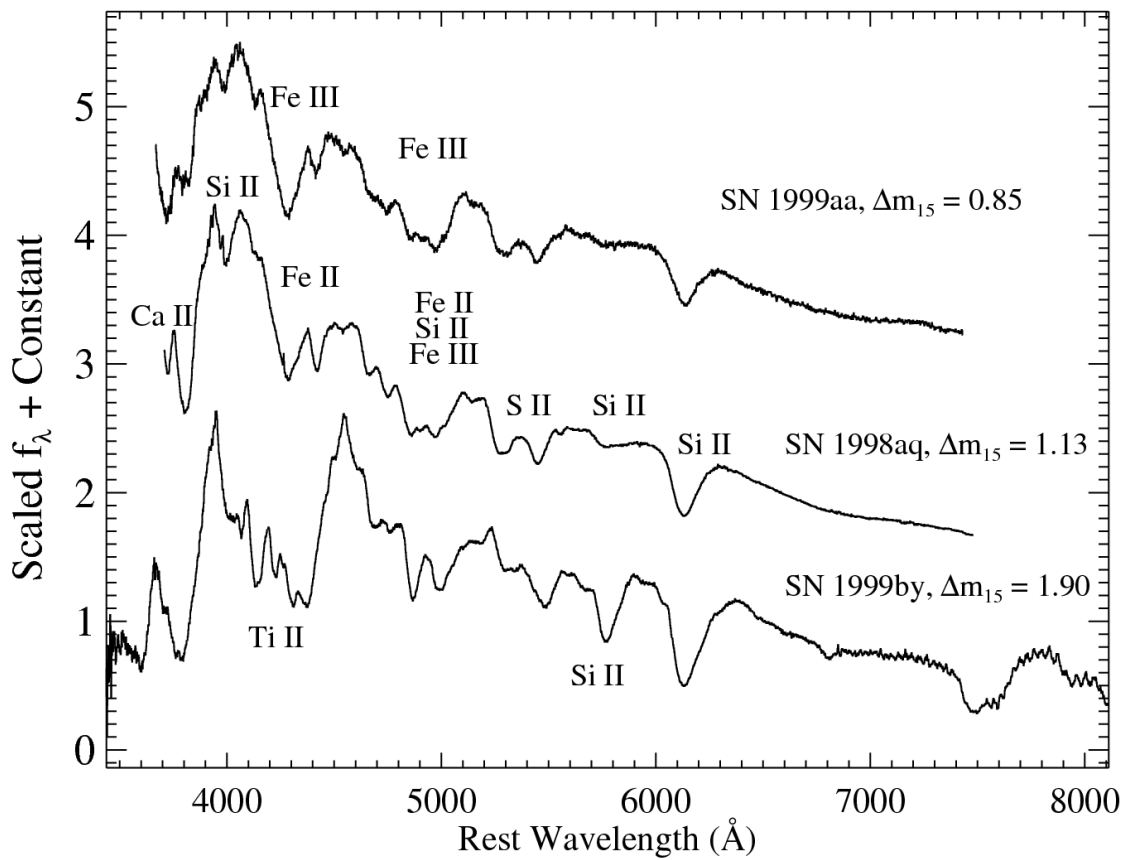}}
\caption{Spectra of SNe~1998aq, 1999aa, and 1999by at $B$-band
  maximum.  The flux units are f$_{\lambda}$ (ergs s$^{-1}$ cm$^{-2}$
  \AA$^{-1}$) that have been normalized and then additive offsets
  applied for clarity. The systemic heliocentric velocity listed in
  Table \ref{sndata} has been removed.  Major features of the spectra
  are labeled.  Note that the overluminous SN~1999aa has weaker
  \ion{Si}{2} than SN~1998aq and stronger \ion{Fe}{3}.  The
  underluminous SN~1999by has stronger \ion{Si}{2} (especially near
  5800 \AA) and \ion{Ti}{2}.\label{dm15}}
\end{figure}

In Figure \ref{labels}, we show two examples of SNe~Ia spectra from
the sample.  SN~1998aq is shown at $B$-band maximum, while SN~2001V is
shown at 20 days past $B$-band maximum.  Both of these SNe are what
would be considered photometrically normal with SN~1998aq having a
\dmm\ of 1.13 \citep{riess05} and SN~2001V having a \dmm\ of 0.99
\citep{mandel07}.  These spectra are fairly typical results for the
brighter SNe.  In addition, we label some major features of the
spectra in order to facilitate discussion of the individual objects
below \citep[see, e.g., ][]{branch05}.  To demonstrate the differences among
the spectra of SNe with different light-curve shapes, we show two
extreme examples compared with the more normal SN~1998aq in Figure
\ref{dm15}.  SN~1999aa has a \dmm\ of 0.85 and was overluminous.  Note
the weaker \ion{Si}{2} and stronger \ion{Fe}{3}.  In contrast,
SN~1999by has a \dmm\ of 1.90 and was subluminous.  It shows stronger
\ion{Si}{2} and \ion{Ti}{2}.

\clearpage

\clearpage

\section{Comments on Individual Supernovae}

\begin{figure}
\plotone{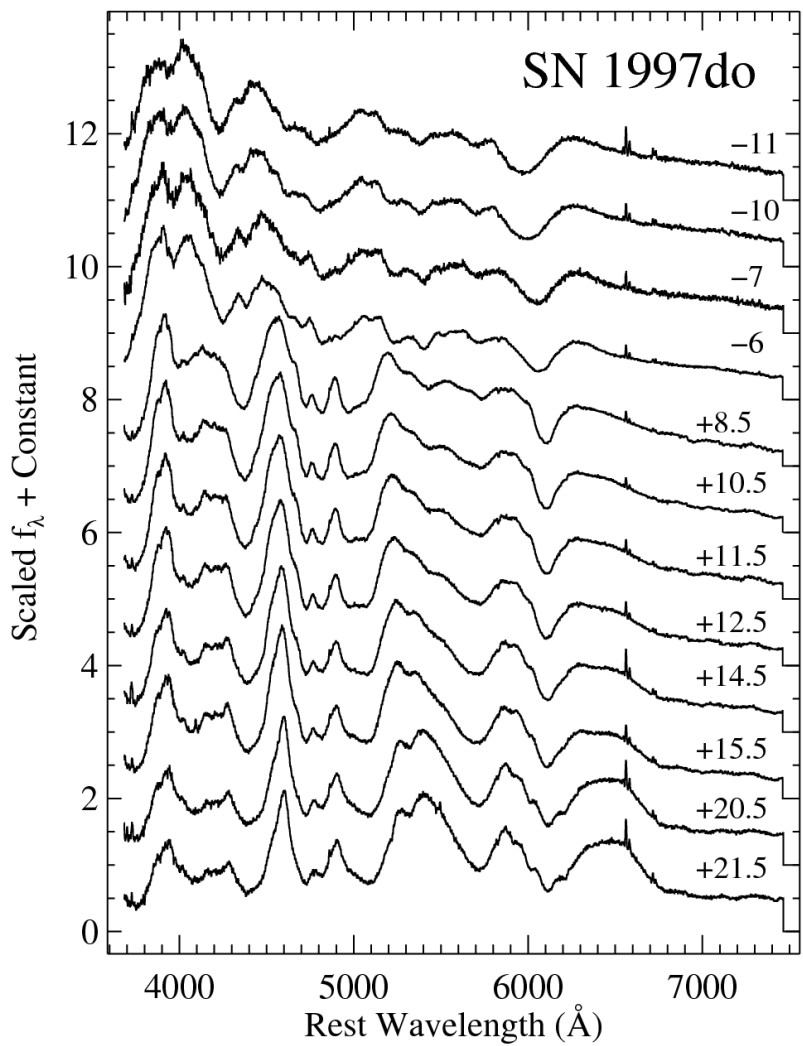}
\caption{Spectra of SN~1997do.  The flux units are f$_{\lambda}$ (ergs
   s$^{-1}$ cm$^{-2}$ \AA$^{-1}$) that have been normalized and then
   additive offsets applied for clarity.  The zero-flux level for each
   spectrum is marked with an extension on the red edge of the
   spectrum (occasionally, this is marked on the blue edge if that produces
   a clearer presentation).  The systemic heliocentric
   velocity listed in Table \ref{sndata} has been
   removed.  The numbers associated with each spectrum indicate the
   epoch of the spectrum relative to $B$-band maximum.\label{97domont}}
\end{figure}

\begin{figure}
\plotone{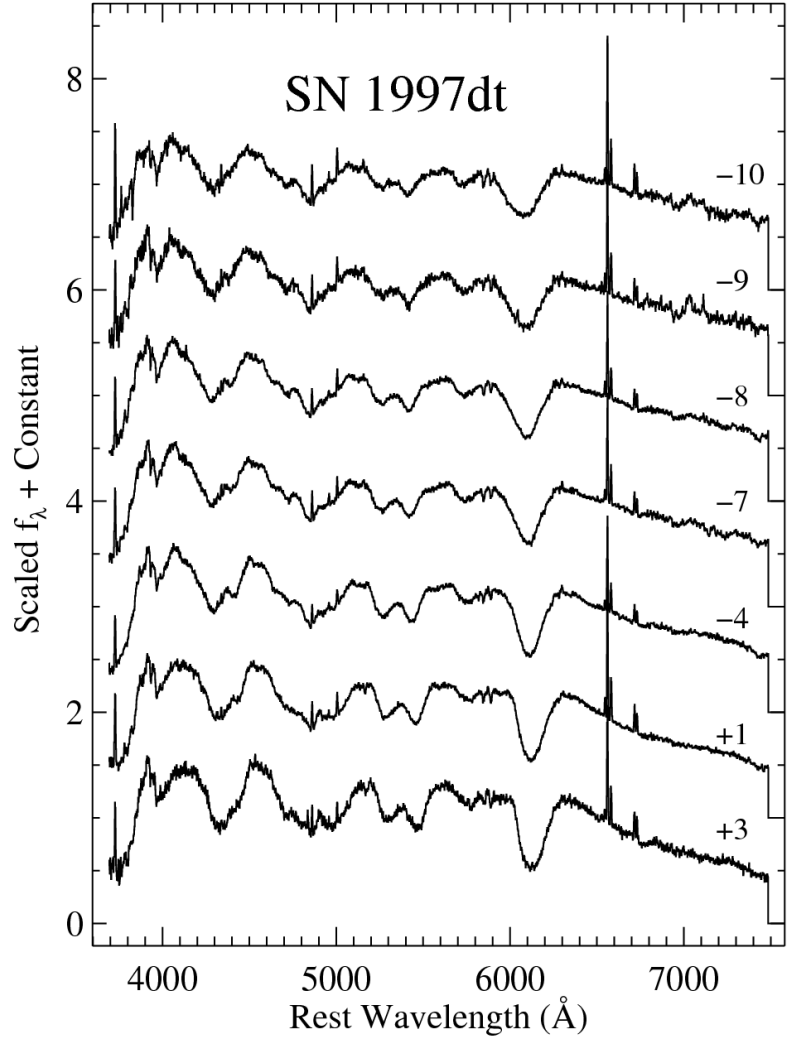}
\caption{Spectra of SN~1997dt.  The flux units, wavelength scale, and
  epoch for each spectrum are as described in Figure
  \ref{97domont}.\label{97dtmont}}
\end{figure}

\begin{figure}
\plotone{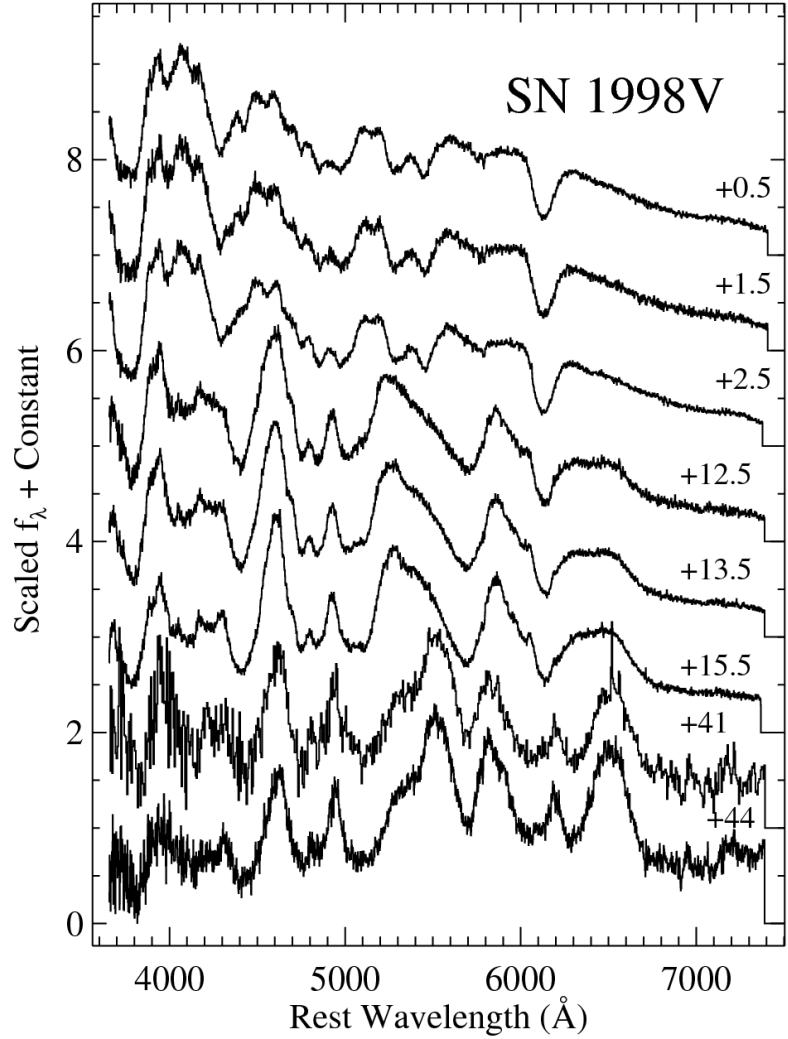}
\caption{Spectra of SN~1998V.  The flux units, wavelength scale, and
  epoch for each spectrum are as described in Figure \ref{97domont}.
  The day +41 spectrum has been rebinned for clarity.\label{98vmont}}
\end{figure}

\emph{SN 1997do}--This SN was discovered on 1997 Oct 31 during the
course of the Beijing Astronomical Observatory (BAO) SN survey
\citep{qiu97}.  A spectrum obtained by Qiu et al. on 1997 Nov 1 showed
that SN~1997do was of Type Ia.  The CfA spectra (Figure
\ref{97domont}) begin eleven days before $B$-band maximum and
continue for three weeks past maximum.
\clearpage
\clearpage
\emph{SN 1997dt}--Another product of the BAO SN survey, this object
was discovered on 1997 Nov 22 \citep{qiao97}.  Qiao et al. also
reported that a spectrum taken the same night as the discovery
indicated that SN~1997dt was of Type Ia.  The CfA spectra (Figure
\ref{97dtmont}) begin ten days before $B$-band maximum, with good
coverage through two days past the time of maximum.  There is some
galaxy contamination in the spectra, as shown by the strong narrow
emission lines apparent in the spectra.
\clearpage
\clearpage
\emph{SN 1998V}--This SN was discovered by the U.K. Nova/Supernova
Patrol on 1998 Mar 10 \citep{hurst98b}.  It was classified as an SN~Ia
\citep{jha98d} based up the first spectrum in the CfA sample (Figure
\ref{98vmont}), obtained at maximum.  We have some coverage of the
post-maximum decline and later phases.
\clearpage
\clearpage

\begin{figure}
\plotone{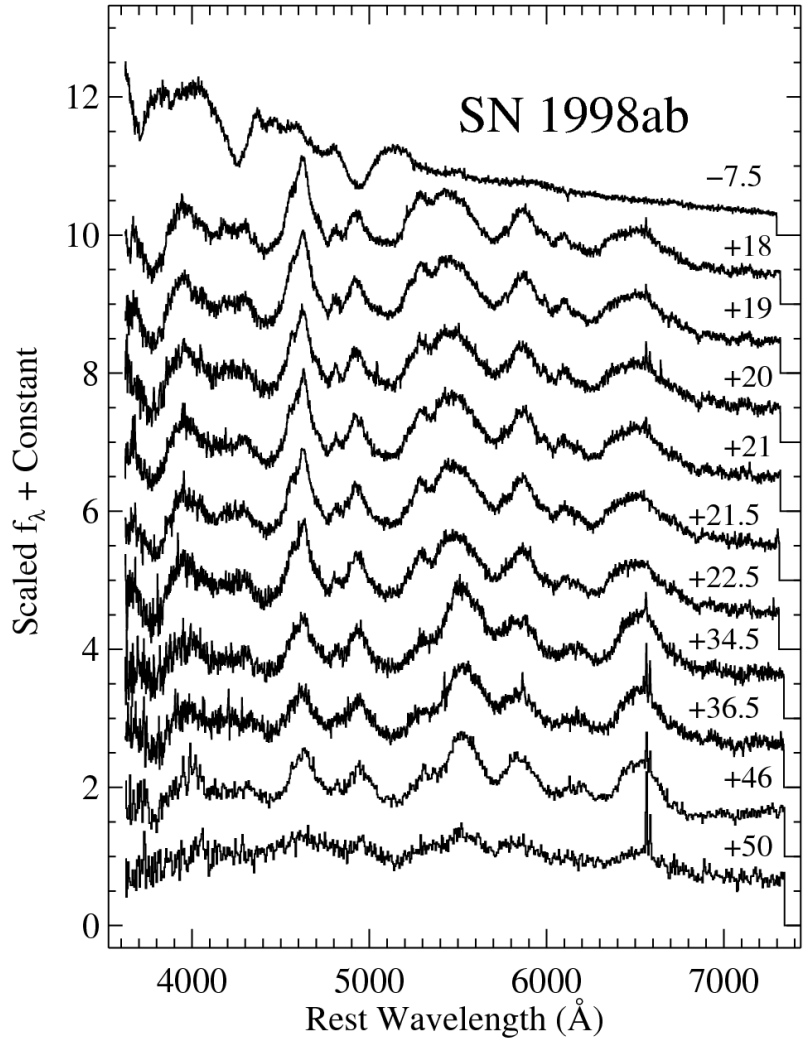}
\caption{Spectra of SN~1998ab.  The flux units, wavelength scale, and
  epoch for each spectrum are as described in Figure \ref{97domont}.
  The days +36.5, +46, and +50 spectra have been rebinned for
  clarity.\label{98abmont}}
\end{figure}

\begin{figure}
\plotone{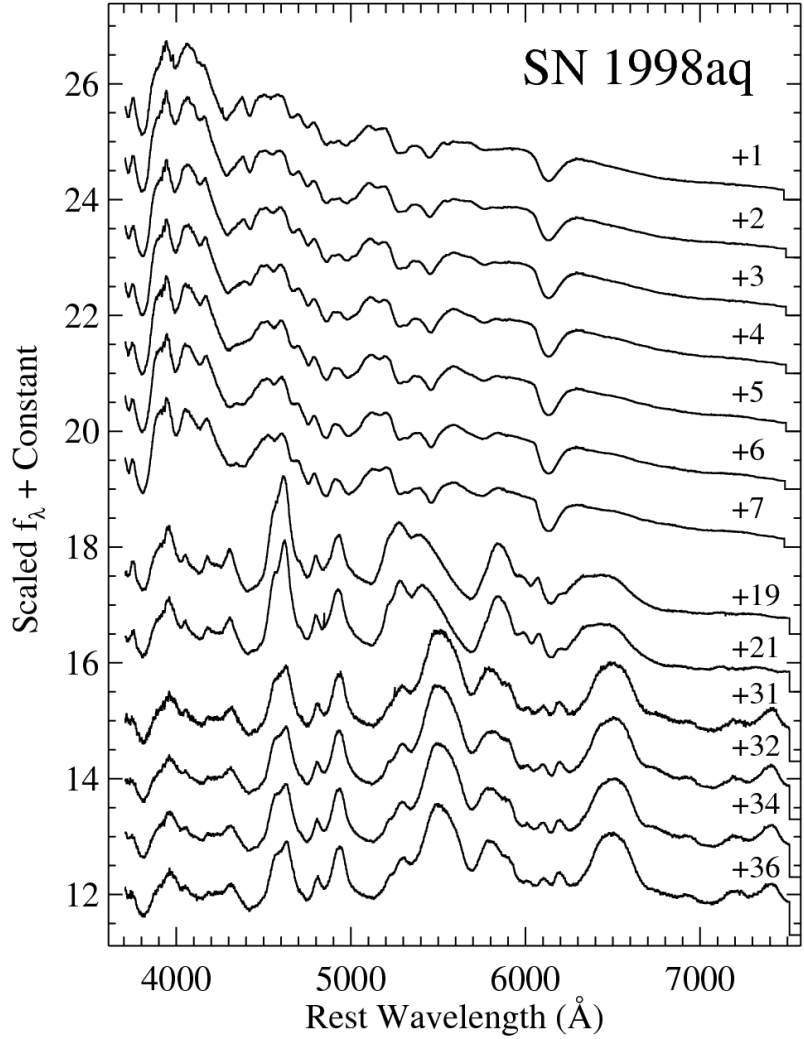}
\caption{Early spectra of SN~1998aq.  The flux units, wavelength
  scale, and epoch for each spectrum are as described in Figure
  \ref{97domont}.  \label{98aqmonta}}
\end{figure}

\begin{figure}
\plotone{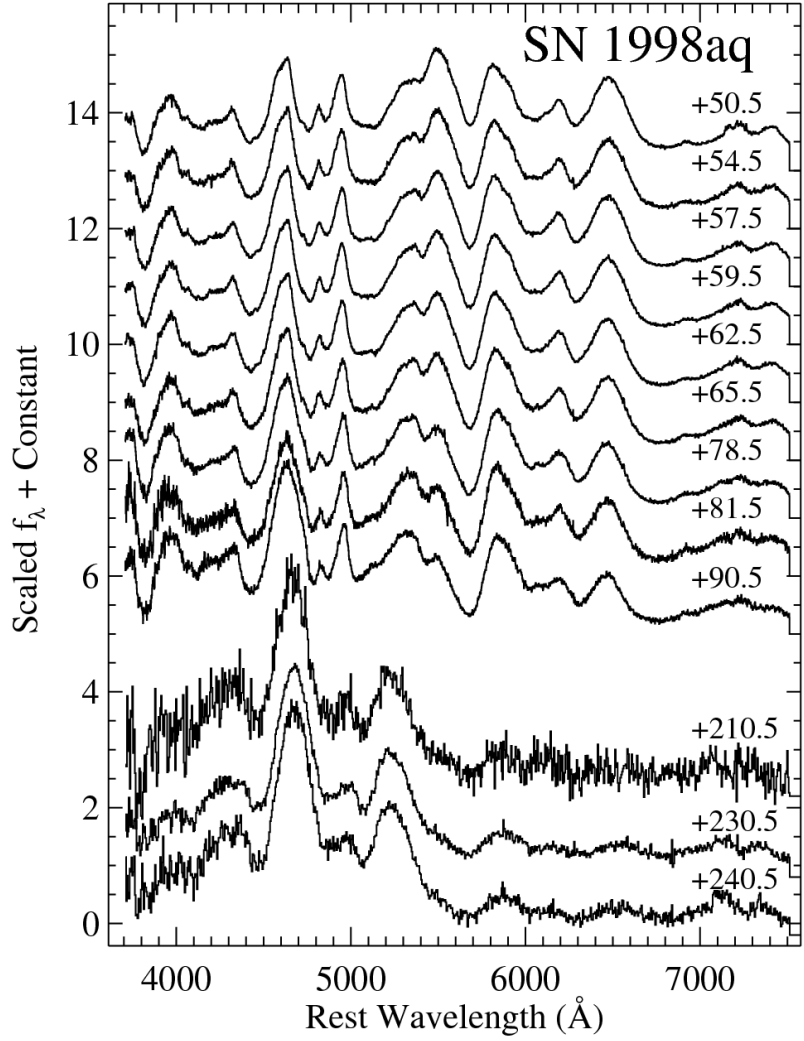}
\caption{Late spectra of SN~1998aq.  The flux units, wavelength scale,
  and epoch for each spectrum are as described in Figure
  \ref{97domont}.  \label{98aqmontb}}
\end{figure}

\emph{SN 1998ab}--The BAO SN survey also found SN~1998ab on 1998 Apr 1
\citep{wei98}.  The first CfA spectrum (Figure \ref{98abmont}) was
used to report the Type of the SN as Ia \citep{garnavich98b}.  This
spectrum was obtained eight days before maximum.  In addition,
Garnavich et al. noted that the \ion{Si}{2} feature was not apparent
in the spectrum, but absorptions associated with \ion{Fe}{3} were
present, indicating that this was a spectroscopically peculiar SN
similar to SN~1991T \citep{filippenko92a, phillips92} at early epochs.
Unfortunately, we were not able to obtain more spectra of this object
in the photospheric phase, but there is a large amount of coverage
at several months past $B$-band maximum.
\clearpage
\clearpage
\emph{SN 1998aq}--Another SN discovered by the U.K. Nova/Supernova
Patrol, SN~1998aq was found on 1998 Apr 13 \citep{hurst98a}.
\citet{ayani98b} and \citet{garnavich98c} reported that the spectrum
showed SN~1998aq to be of Type Ia.  Ayani \& Yamaoka also felt that
the strength of the \ion{Si}{2} $\lambda$5800 line might indicate that
SN~1998aq was subluminous, but the full set of CfA spectra (Figures
\ref{98aqmonta} and \ref{98aqmontb}) shows that it was
spectroscopically normal.  Although there are no pre-maximum spectra,
there is good coverage starting with the first spectrum taken one day
past maximum, continuing with almost daily spectra in the weeks past
maximum.  There is an extensive set of spectra obtained several weeks
past maximum, as well as a few in the nebular phase.  Some of these
spectra have been published and analyzed by \citet{branch03}, but we
include them here for completeness.
\clearpage

\begin{figure}
\plotone{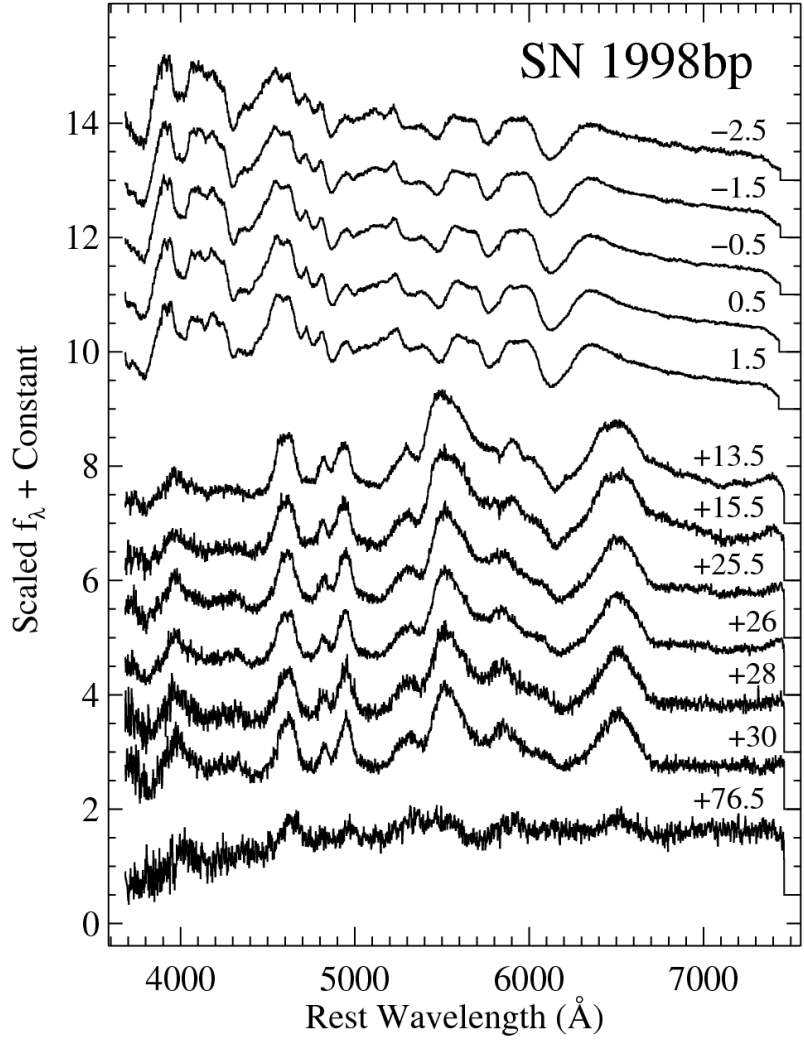}
\caption{Spectra of SN~1998bp.  The flux units, wavelength scale, and
  epoch for each spectrum are as described in Figure
  \ref{97domont}.\label{98bpmont}}
\end{figure}

\clearpage
\emph{SN 1998bp}--This SN was also found by the U.K. Nova/Supernova
Patrol on 1998 Apr 29 \citep{hurst98d}.  A spectrum taken the next
night showed that is was of Type Ia, but possibly spectroscopically
peculiar \citep{patat98}.  The first CfA spectrum (Figure
\ref{98bpmont}, obtained three days before maximum) also prompted a
report that the object seemed peculiar \citep{jha98b}, specifically
that the \ion{Si}{2} $\lambda$5800 line was strong.  As seen in Figure
\ref{98bpmont}, the strength of the $\lambda$5800 feature is similar
to what is observed in subluminous SNe~Ia \citep[e.g., SN~1991bg
][]{filippenko92b, leibundgut93}, but the blue half of the spectrum
appears more normal.  The \dmm value of 1.83 from \citet{jha06}
confirms that this was a peculiar SN~Ia.  The CfA sample has good
coverage near maximum, as well as several at a few weeks past maximum.
\clearpage
\clearpage

\begin{figure}
\plotone{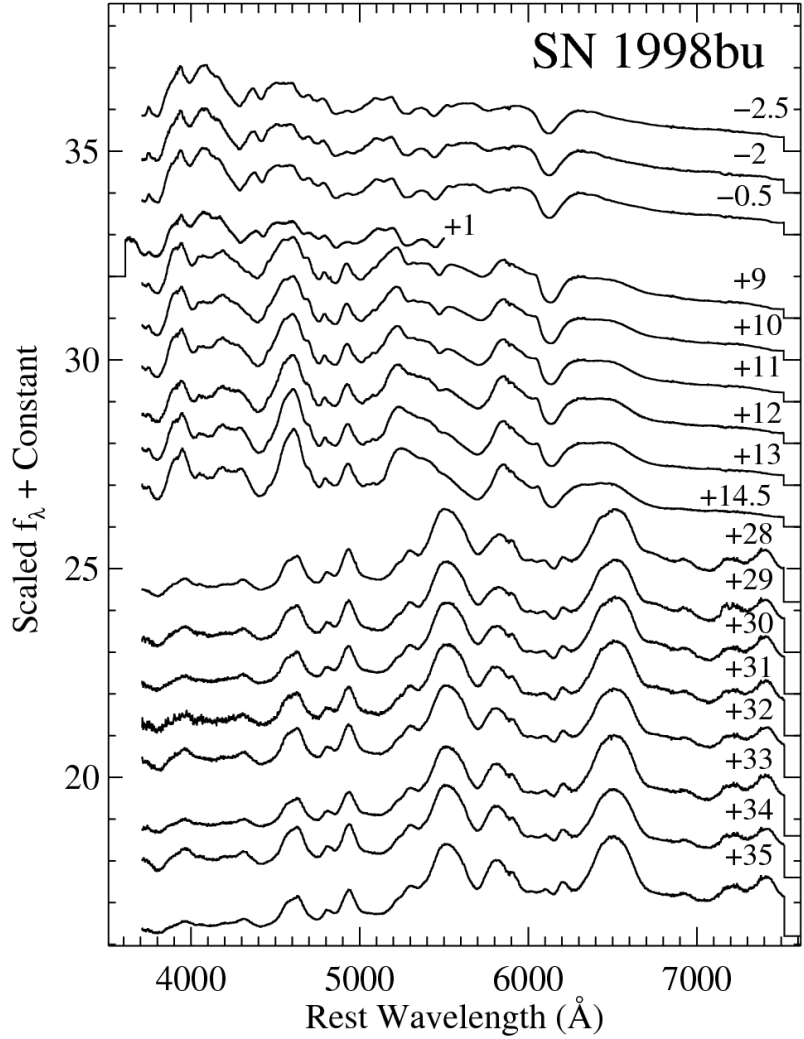}
\caption{Early spectra of SN~1998bu.  The flux units, wavelength
  scale, and epoch for each spectrum are as described in Figure
  \ref{97domont}.  \label{98bumonta}}
\end{figure}

\begin{figure}
\plotone{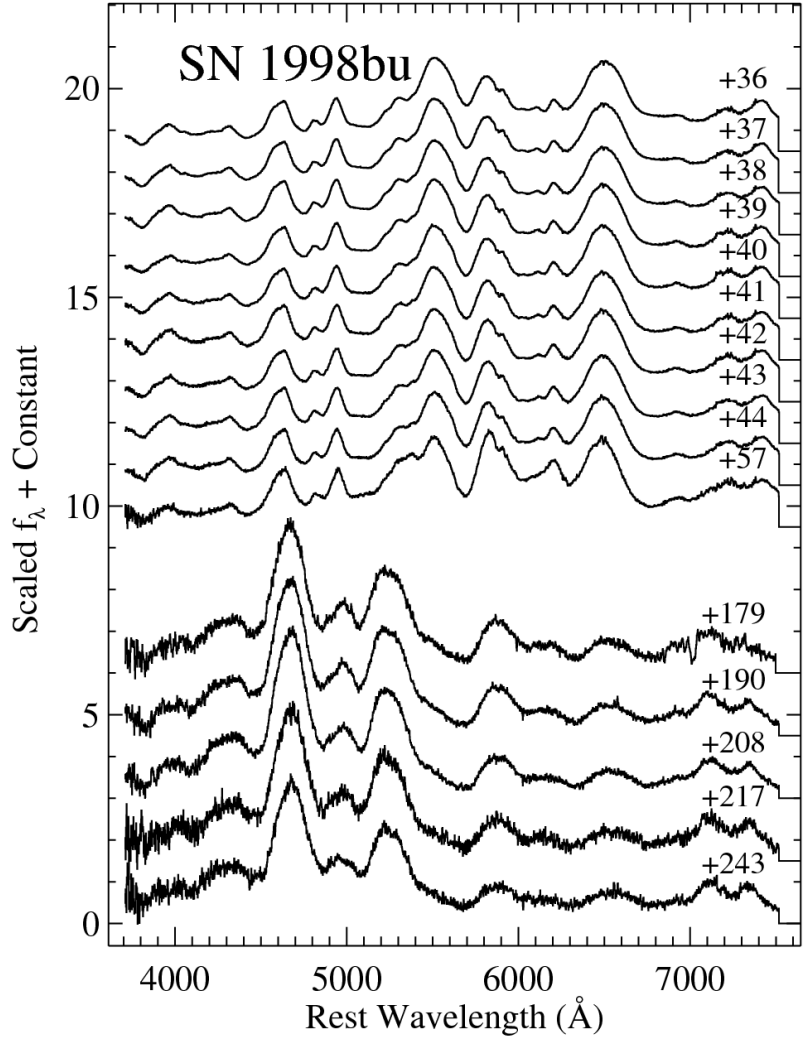}
\caption{Late spectra of SN~1998bu.  The flux units, wavelength scale,
  and epoch for each spectrum are as described in Figure
  \ref{97domont}.  \label{98bumontb}}
\end{figure}

\emph{SN 1998bu}--This SN was discovered by \citet{villi98} on 1998
May 9.  Two groups reported that spectroscopy indicated that SN~1998bu
was of Type Ia \citep{ayani98a}.  The CfA sample of spectra is large
(Figures \ref{98bumonta} and \ref{98bumontb}), with good coverage near
maximum (the first being three days before maximum), as well as many
spectra up to two months past maximum and few in the nebular phase.
Some of the CfA spectra were shown and discussed by \citet{jha99b},
but the spectra presented herein were rereduced to be consistent with
the rest of this spectroscopic sample.  This was a bright SN that was
followed extensively by many groups \citep[e.g., ][]{suntzeff99}.
\clearpage

\begin{figure}
\plotone{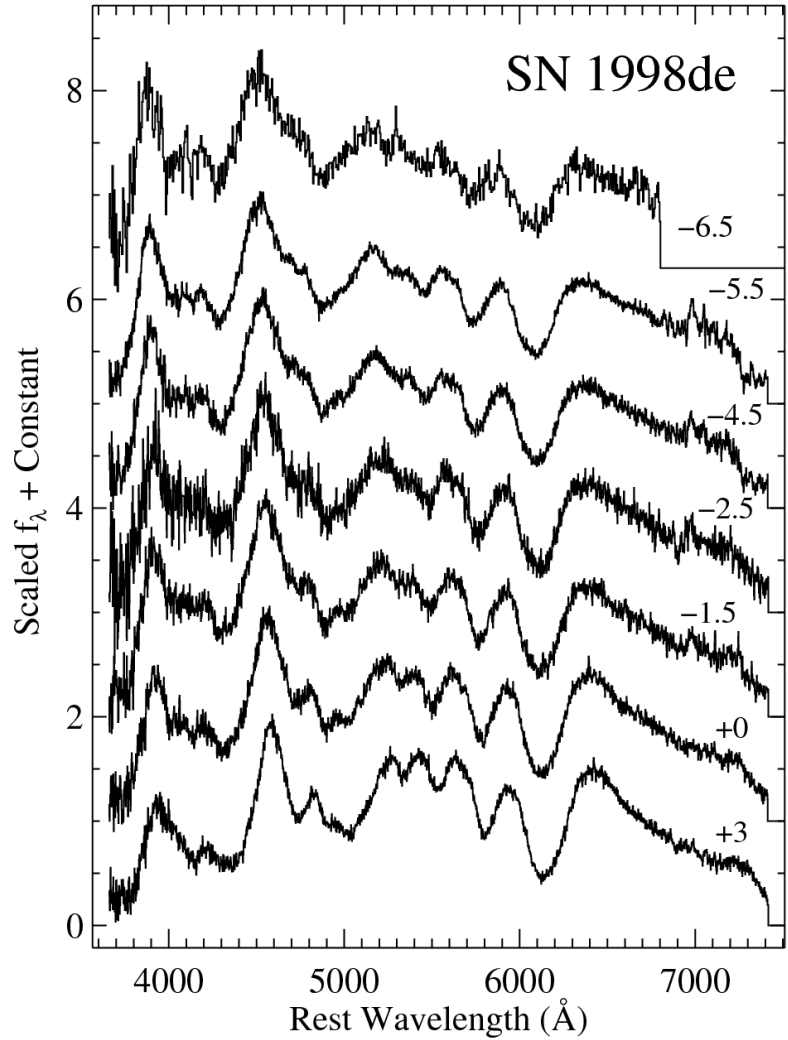}
\caption{Spectra of SN~1998de.  The flux units, wavelength scale, and
  epoch for each spectrum are as described in Figure \ref{97domont}.
  The day -6.5 spectrum has been trimmed and rebinned for
  clarity.\label{98demont}}
\end{figure}

\clearpage
\emph{SN 1998de}--This SN was found in the course of the Lick
Observatory Supernova Search (LOSS) on 1998 Jul 23 \citep{modjaz98a}.
The initial classification as a Type Ia was based upon our first CfA
spectrum (Figure \ref{98demont}) obtained seven days before maximum.
This SN also appeared to be spectroscopically peculiar, with a strong
\ion{Si}{2} $\lambda$5800 line and prominent \ion{Ti}{2} absorptions
in the blue \citep{garnavich98d}.  This characterization as a
subluminous SN~Ia was confirmed photometrically and spectroscopically
by \citet{modjaz01} \citep[and the \dmm value of 1.93 from ][]{jha06}
and can be seen in the series of spectra of Figure
\ref{98demont}.
\clearpage
\clearpage

\begin{figure}
\plotone{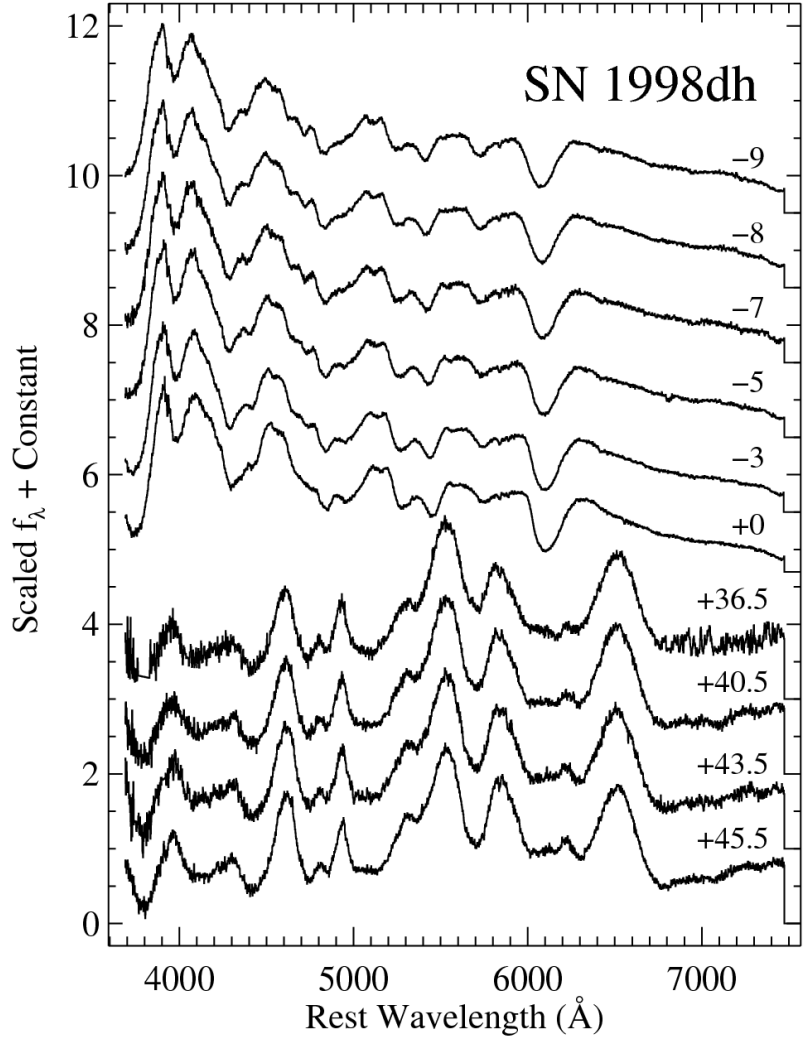}
\caption{Spectra of SN~1998dh.  The flux units, wavelength scale, and
  epoch for each spectrum are as described in Figure
  \ref{97domont}.\label{98dhmont}}
\end{figure}

\emph{SN 1998dh}-- Another product of the LOSS, SN~1998dh was found on
1998 Jul 20 \citep{li98}.  The first CfA spectrum (Figure
\ref{98dhmont}), obtained nine days before maximum, was used to
report that it was an SN~Ia \citep{garnavich98a}.  There is good
coverage before maximum, as well as some later coverage.
\clearpage

\begin{figure}
\plotone{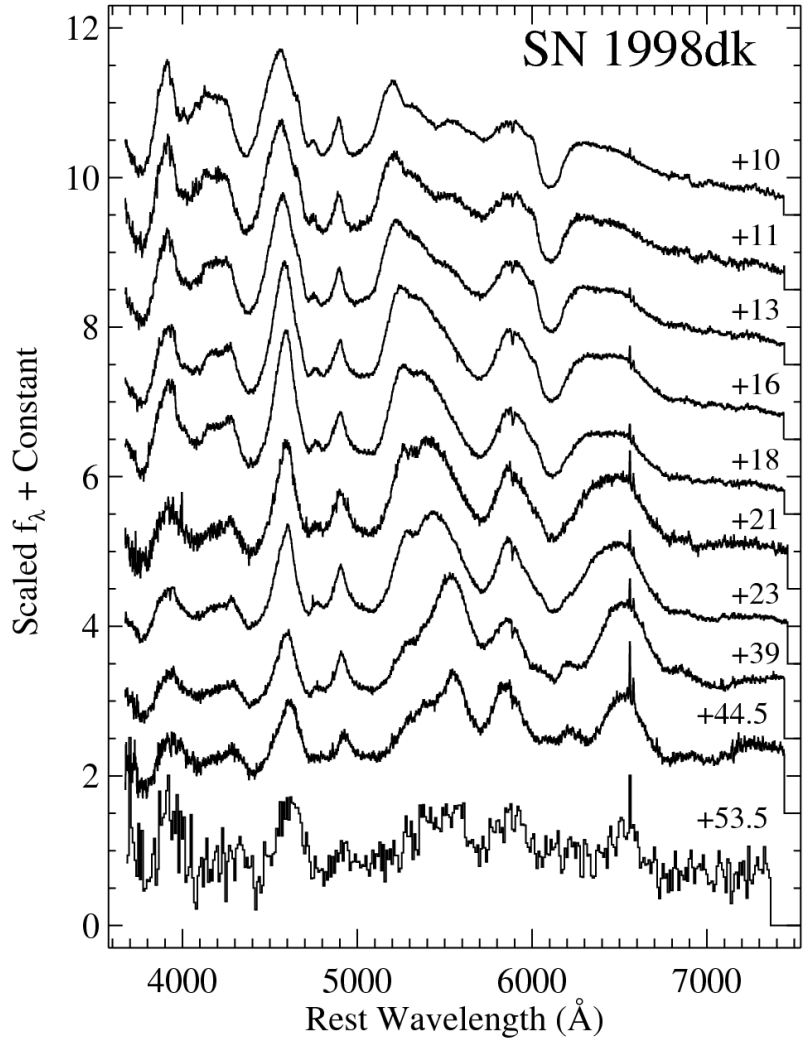}
\caption{Spectra of SN~1998dk.  The flux units, wavelength scale, and
  epoch for each spectrum are as described in Figure \ref{97domont}.
  The day +53.5 spectrum has been rebinned for clarity.\label{98dkmont}}
\end{figure}

\begin{figure}
\plotone{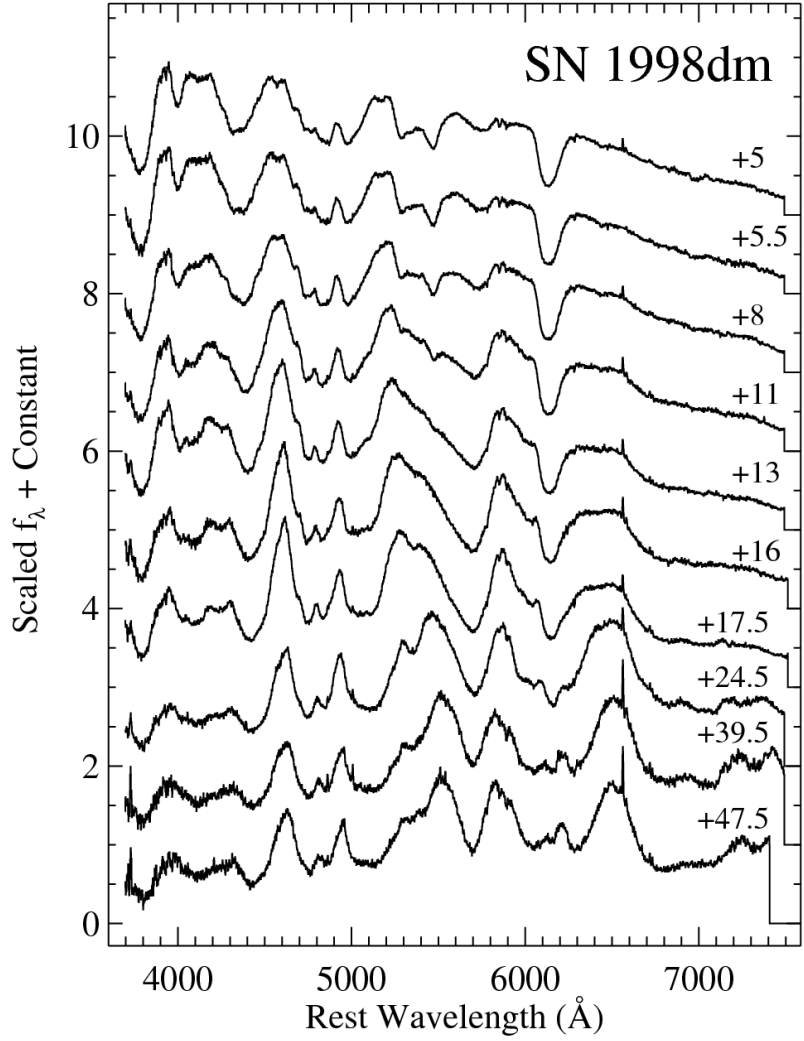}
\caption{Spectra of SN~1998dm.  The flux units, wavelength scale, and
  epoch for each spectrum are as described in Figure
  \ref{97domont}.\label{98dmmont}}
\end{figure}

\begin{figure}
\plotone{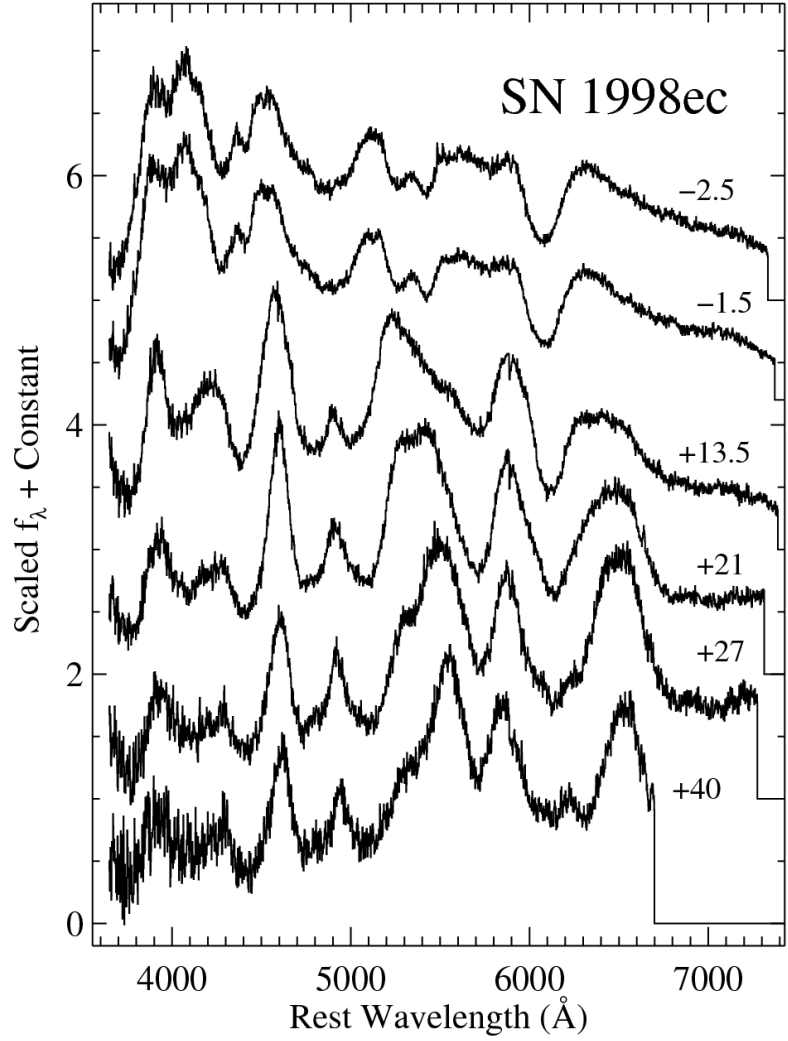}
\caption{Spectra of SN~1998ec.  The flux units, wavelength scale, and
  epoch for each spectrum are as described in Figure \ref{97domont}.
  The day +40 spectrum has been trimmed for clarity.\label{98ecmont}}
\end{figure}

\begin{figure}
\plotone{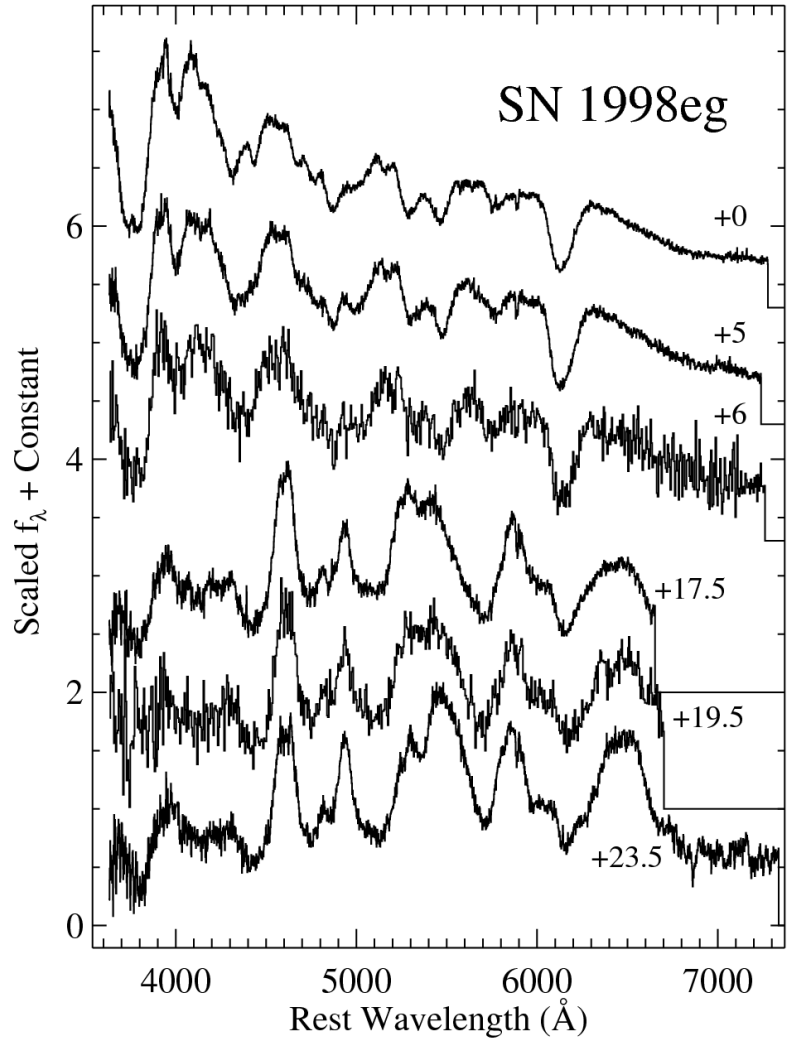}
\caption{Spectra of SN~1998eg.  The flux units, wavelength scale, and
  epoch for each spectrum are as described in Figure \ref{97domont}.
  The days +5 and +19.5 spectra have been rebinned and the days +17.5 and
  +19.5 spectra have been trimmed for clarity.\label{98egmont}}
\end{figure}

\begin{figure}
\plotone{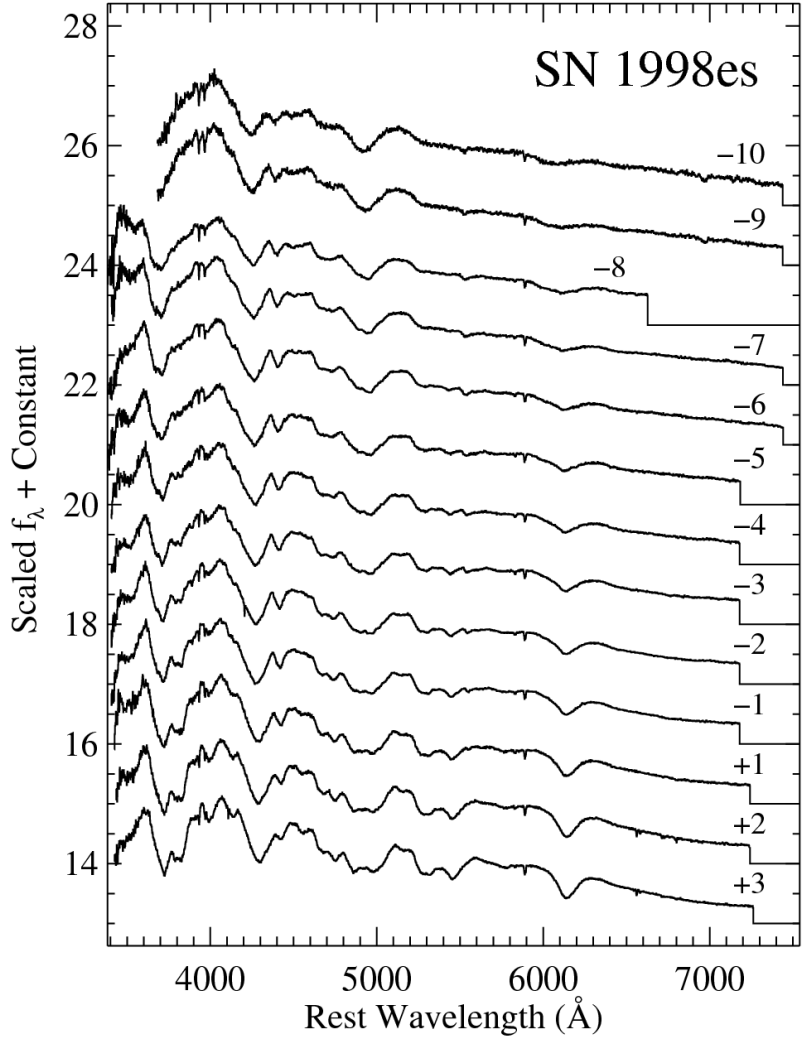}
\caption{Early spectra of SN~1998es.  The flux units, wavelength
  scale, and epoch for each spectrum are as described in Figure
  \ref{97domont}.\label{98esmonta}}
\end{figure}

\begin{figure}
\plotone{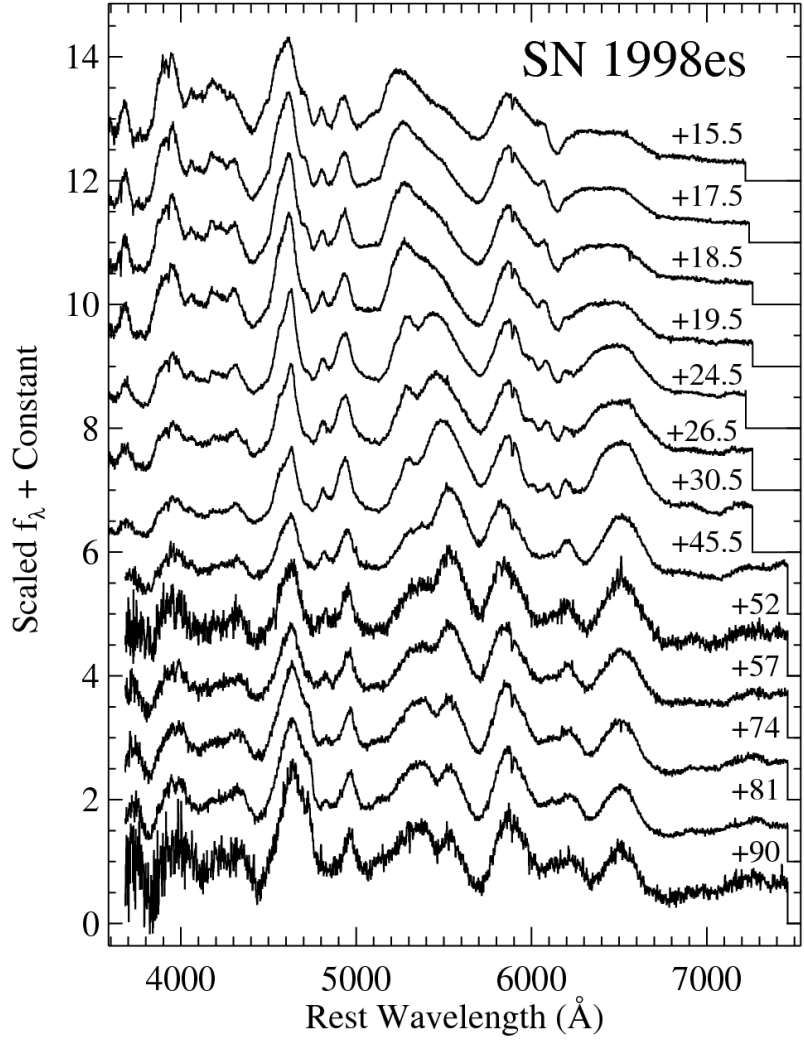}
\caption{Late spectra of SN~1998es.  The flux units, wavelength scale,
  and epoch for each spectrum are as described in Figure
  \ref{97domont}.\label{98esmontb}}
\end{figure}

\clearpage
\emph{SN 1998dk}--The LOSS also discovered SN~1998dk on 1998 Aug
19 \citep{king98}.  It was classified as an SN Ia \citep{filippenko98}.
The telescopes on Mt. Hopkins are traditionally closed during August
for the Arizona monsoon season, so the CfA spectra begin ten days after
maximum (Figure \ref{98dkmont}), but with several spectra in the weeks
past maximum.
\clearpage
\clearpage
\emph{SN 1998dm}--This SN was also found by LOSS on 1998 Aug 22
\citep{modjaz98b} and SN~1998dm was subsequently classified as an
SN~Ia \citep{filippenko98}.  Filippenko \& De Breuck noted that the
spectra were unusually red and that the \ion{Si}{2} $\lambda$5800 line
seemed relatively strong, suggesting that this was a subluminous
event.  Our spectra (Figure \ref{98dmmont}) appear more normal
(although our early coverage was limited by the August shutdown
described above, leading to the earliest CfA spectrum being five days
past maximum), and the \dmm\ value of 1.07 \citep{jha06} confirms that
SN~1998dm was not clearly peculiar.
\clearpage
\clearpage
\emph{SN 1998ec}--The BAO supernova survey found SN~1998ec on 1998 Sep
26 \citep{qiu98}.  Our first CfA spectrum (Figure \ref{98ecmont}),
observed three days before maximum, was used to classify it as an SN~Ia
\citep{jha98a}.  This object has a more limited number of spectra, but
the pre-maximum observations are valuable.
\clearpage
\clearpage
\emph{SN 1998eg}--Another product of the U.K. Supernova Patrol,
SN~1998eg was found on 1998 Oct 19 \citep{hurst98c}.  Two groups
(including the CfA SN group using our first spectrum obtained at
maximum) separately classified it as an SN~Ia \citep{salvo98}.  Our
spectra cover the period from maximum to a few weeks past maximum
(Figure \ref{98egmont}).
\clearpage
\clearpage
\emph{SN 1998es}--The LOSS discovered SN~1998es on 1998 Nov 13
\citep{halderson98}.  Our first CfA spectrum (Figure \ref{98esmonta})
was used to classify it as an SN~Ia \citep{jha98c}.  Jha et al. noted
that the \ion{Si}{2} $\lambda$6355 feature was relatively weak and
that the spectrum overall resembled that of SN~1991T
\citep{filippenko92a, phillips92} at early epochs.  The
\dmm\ value of 0.87 \citep{jha06} also indicates that this
was a 91T-like event.  The CfA sample (Figures \ref{98esmonta} and
\ref{98esmontb}) begins at ten days before maximum,
and continue with good coverage until just past maximum.  There is
also a large set of spectra covering several few weeks past maximum.
\clearpage

\begin{figure}
\plotone{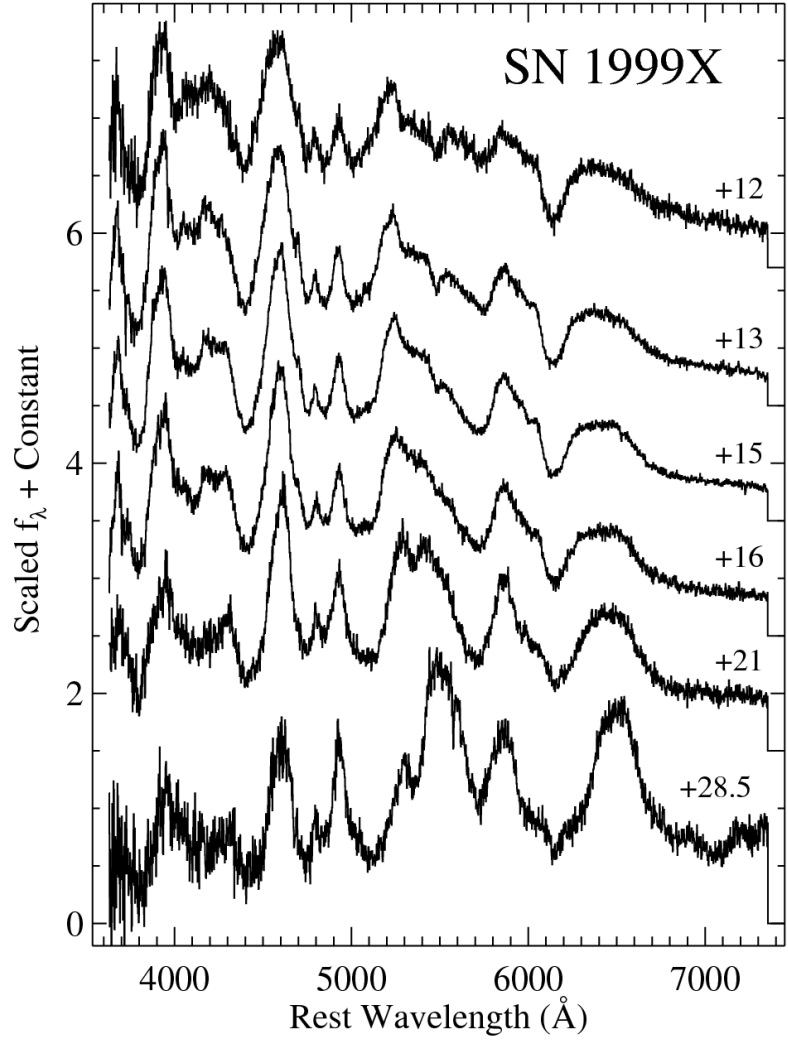}
\caption{Spectra of SN~1999X.  The flux units, wavelength scale, and
  epoch for each spectrum are as described in Figure
  \ref{97domont}.\label{99xmont}}
\end{figure}

\begin{figure}
\plotone{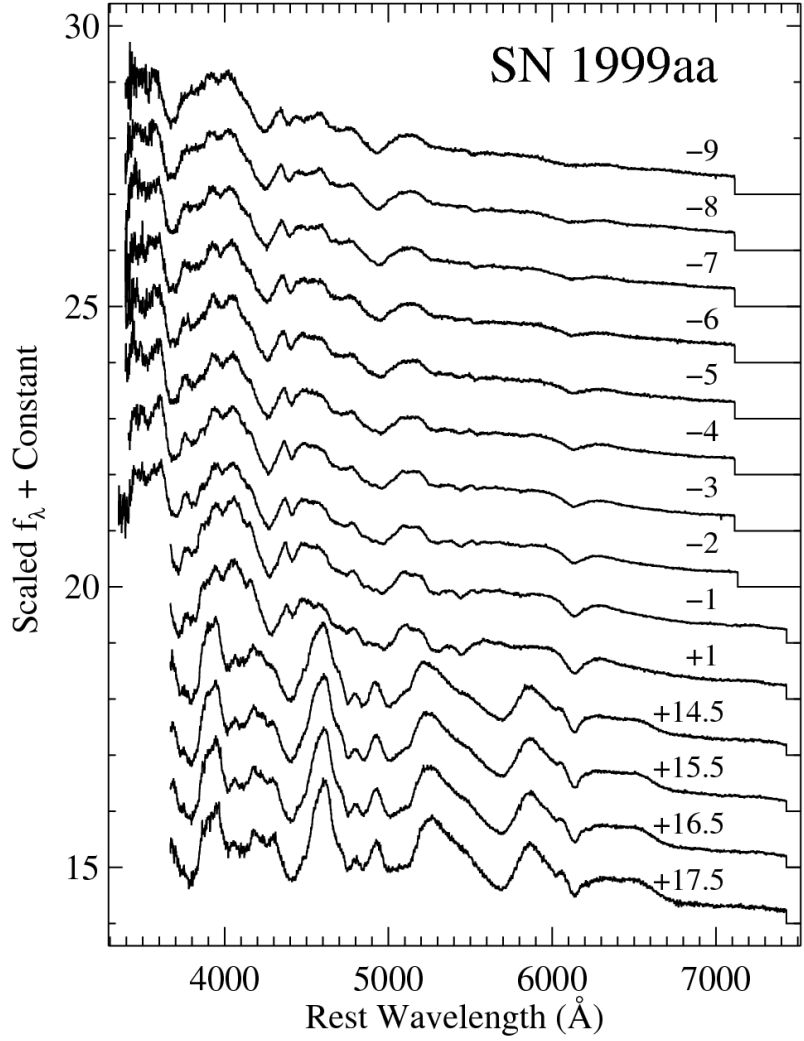}
\caption{Early spectra of SN~1999aa.  The flux units, wavelength
  scale, and epoch for each spectrum are as described in Figure
  \ref{97domont}.\label{99aamonta}}
\end{figure}

\begin{figure}
\plotone{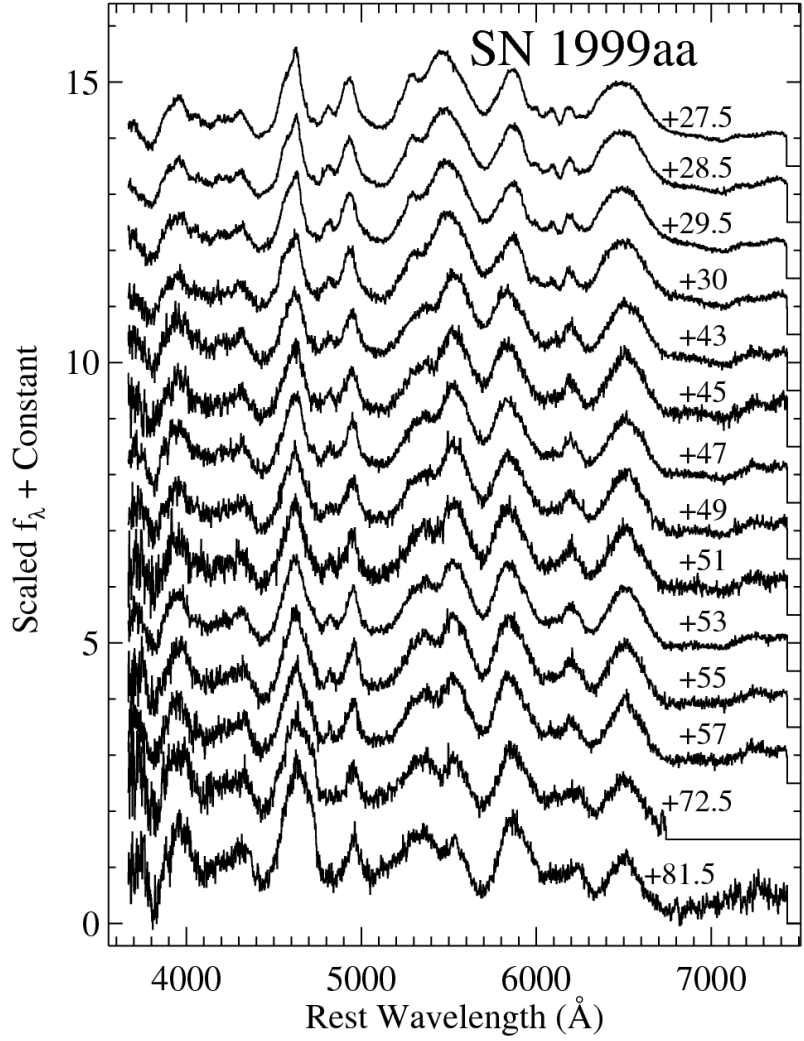}
\caption{Late spectra of SN~1999aa.  The flux units, wavelength scale,
  and epoch for each spectrum are as described in Figure
  \ref{97domont}.\label{99aamontb}}
\end{figure}

\clearpage
\emph{SN 1999X}--This SN was found by \citet{schwartz99} on 1999 Jan
27.  Our first CfA spectrum (Figure \ref{99xmont}), obtained twelve
days after maximum, was used to classify SN~1999X as an SN~Ia
\citep{garnavich99a}.  The spectra cover two to four weeks past
maximum.
\clearpage
\clearpage
\emph{SN 1999aa}--\citet{arbour99a} found SN~1999aa on 1999 Feb 11
(independent discoveries were also reported by \citet{qiao99} and
\citet{nakano99b}).  A spectrum obtained the next day revealed that
SN~1999aa was a spectroscopically peculiar SN~Ia
\citep{filippenko99a}.  The spectrum had some similarities to SN~1991T
\citep{filippenko92a, phillips92}, with weak \ion{Si}{2} $\lambda$6355
and absorptions due to \ion{Fe}{3}.  One difference was an absorption
near 3750 \AA.  This feature (\ion{Ca}{2}~H~\&~K) was not present in
SN~1991T.  The \dmm\ value of 0.85 \citep{jha06} shows that it is
similar to SN~1991T photometrically as well.  The CfA spectra have
extensive coverage from nine days before maximum through three months
past maximum (Figures \ref{99aamonta} and \ref{99aamontb}).  This
bright and unusual SN was well observed by many groups \citep[e.g.,
][]{li01b, garavini04}.
\clearpage

\begin{figure}
\plotone{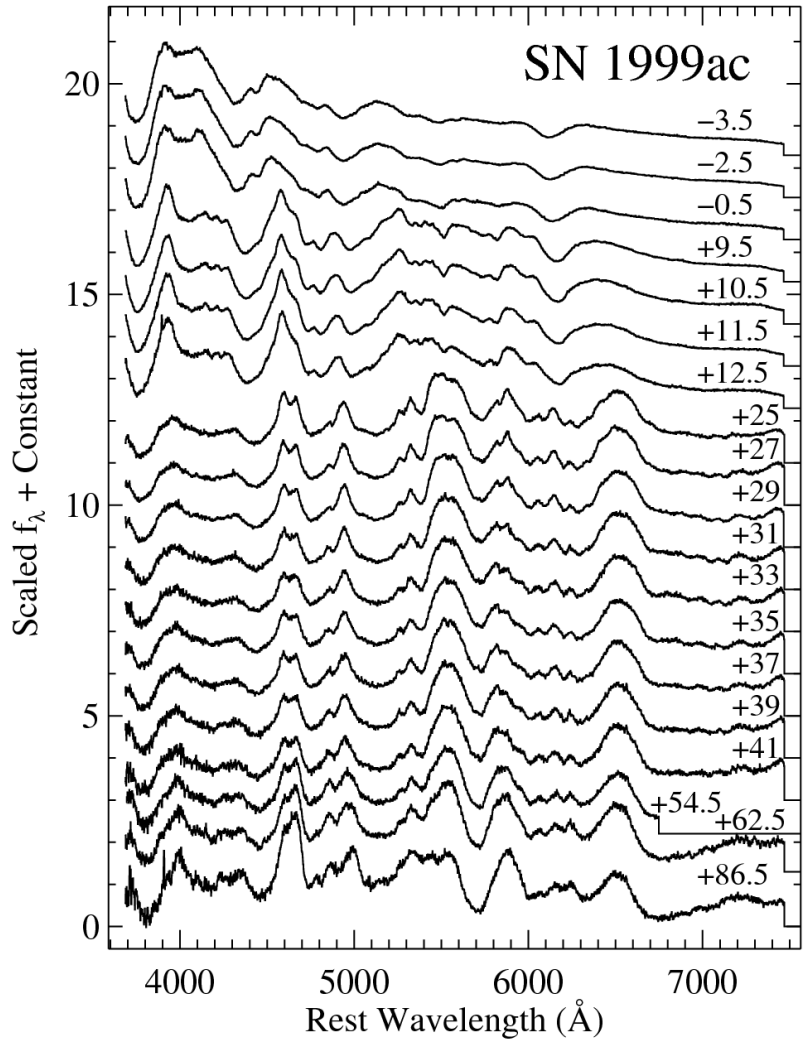}
\caption{Spectra of SN~1999ac.  The flux units, wavelength scale, and
  epoch for each spectrum are as described in Figure
  \ref{97domont}.\label{99acmont}}
\end{figure}

\begin{figure}
\plotone{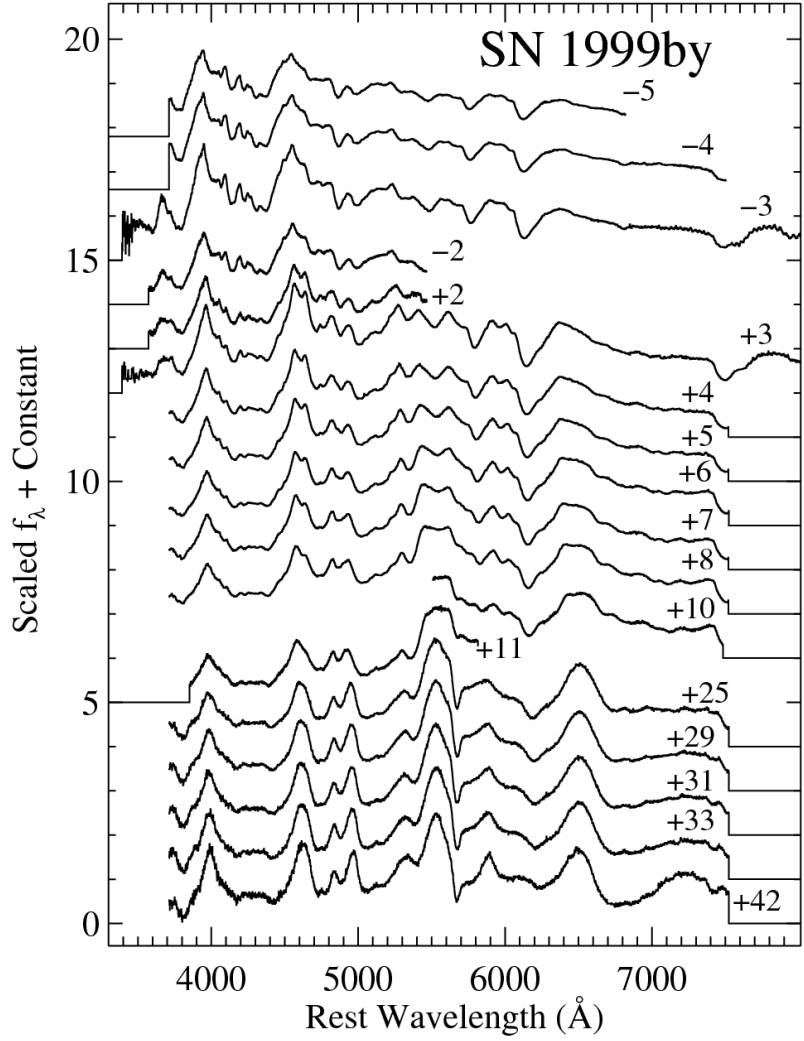}
\caption{Spectra of SN~1999by.  The flux units, wavelength scale, and
  epoch for each spectrum are as described in Figure
  \ref{97domont}.\label{99bymont}}
\end{figure}

\begin{figure}
\plotone{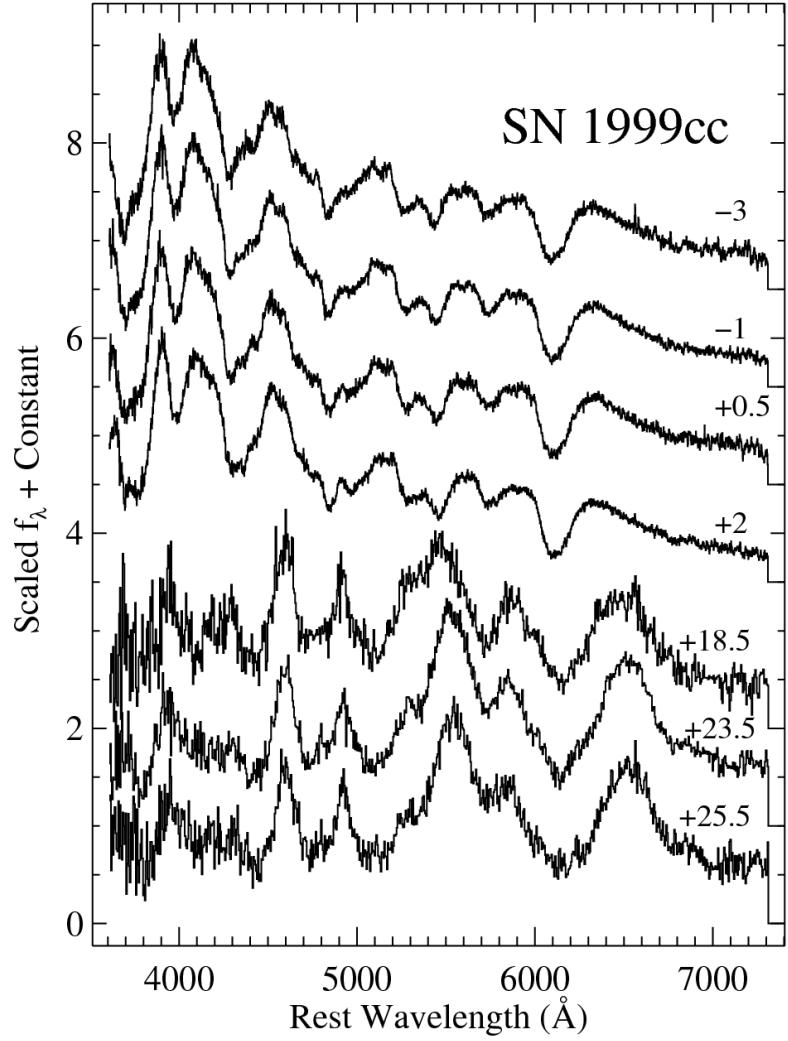}
\caption{Spectra of SN~1999cc.  The flux units, wavelength scale, and
  epoch for each spectrum are as described in Figure
  \ref{97domont}.\label{99ccmont}}
\end{figure}

\begin{figure}
\plotone{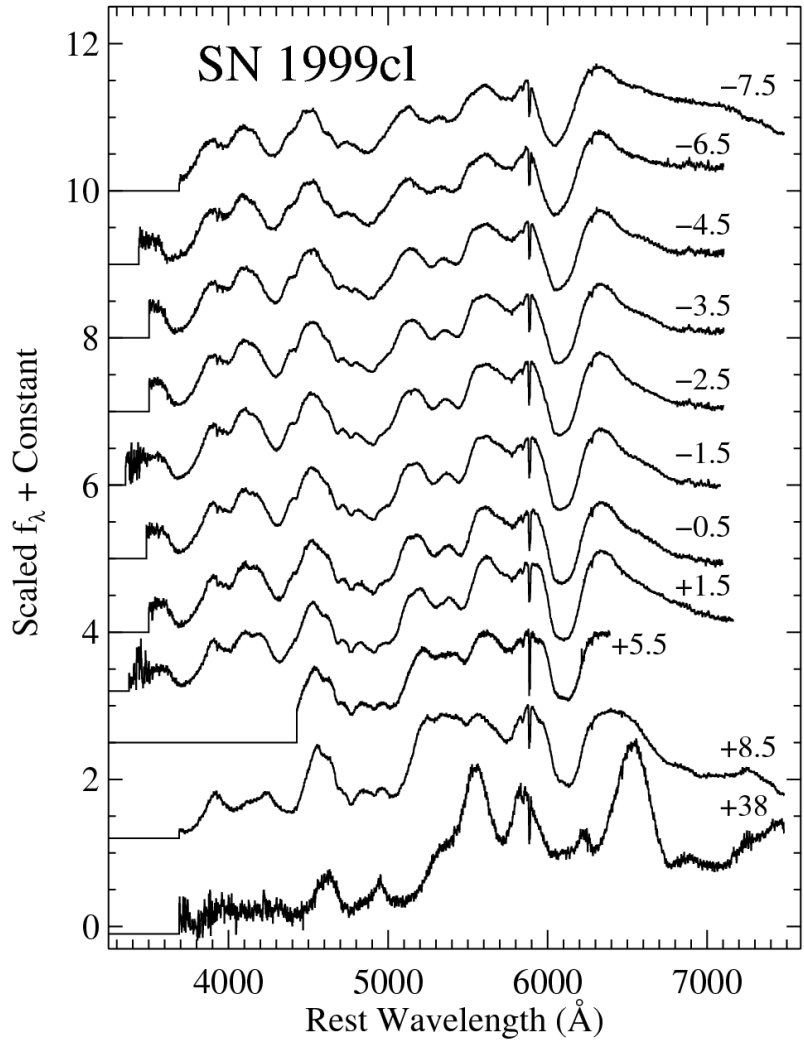}
\caption{Spectra of SN~1999cl.  The flux units, wavelength scale, and
  epoch for each spectrum are as described in Figure
  \ref{97domont}.\label{99clmont}}
\end{figure}

\begin{figure}
\plotone{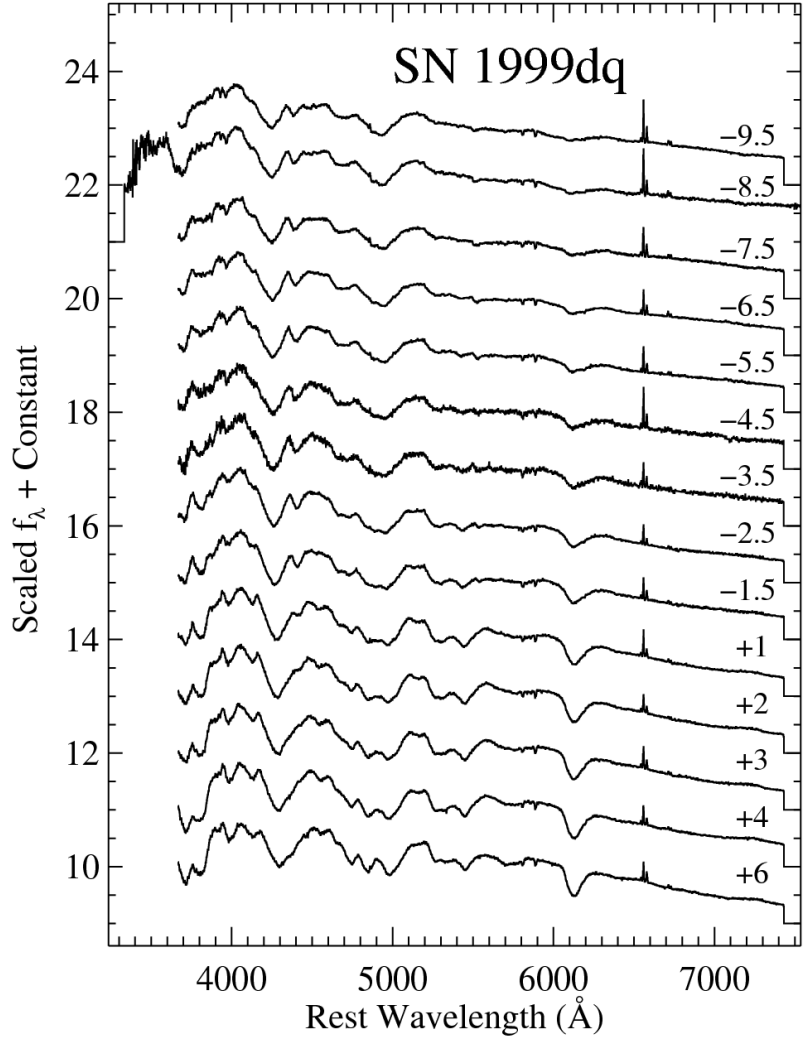}
\caption{Early spectra of SN~1999dq.  The flux units, wavelength
  scale, and epoch for each spectrum are as described in Figure
  \ref{97domont}.\label{99dqmonta}}
\end{figure}

\begin{figure}
\plotone{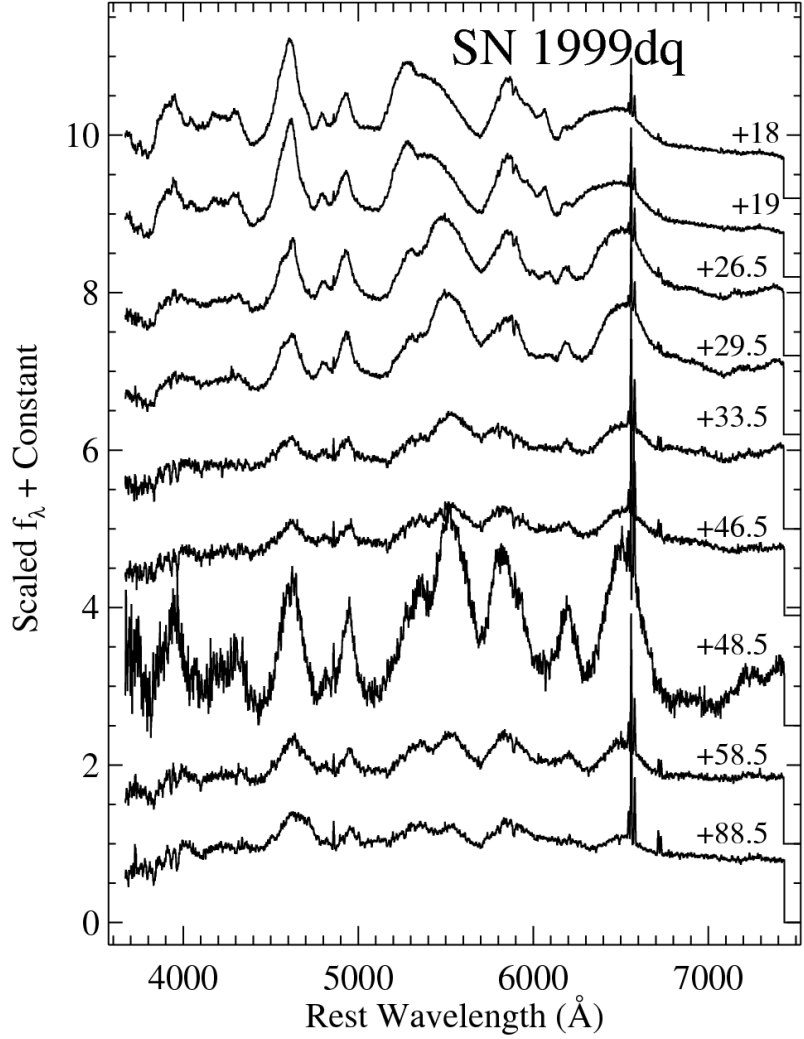}
\caption{Late spectra of SN~1999dq.  The flux units and wavelength
  scale are as described in Figure \ref{97domont}.  Note that for the
  days +33.5, +46.5, +48.5, +58.5, and +88.5 spectra, the relative
  strengths of the lines are dependent on galaxy contamination of the
  spectra.  This can be affected by variable seeing and different
  observed position angles.\label{99dqmontb}}
\end{figure}

\clearpage
\emph{SN 1999ac}--This SN was found by the LOSS on 1999 Feb 26
\citep{modjaz99}.  A spectrum obtained by \citet{phillips99} showed
that it was strikingly similar to SN~1999aa.  The CfA spectra of
SN~1999ac have some near maximum (the first at four days before
maximum) and a large number one to three months past maximum (Figure
\ref{99acmont}).
\clearpage
\clearpage
\emph{SN 1999by}--Both the LOSS and the U.K. Supernova Patrol
independently discovered SN~1999by on 1999 Apr 30 \citep{arbour99b}.
\citet{gerardy99} classified it as an SN~Ia and \citet{garnavich99b}
reported that the spectra showed some signs of peculiarity.  The
\ion{Si}{2} $\lambda$5800 line was relatively strong and \ion{Ti}{2}
features were apparent indicating that this was a subluminous SN~Ia.
This was confirmed photometrically \citep{garnavich04} with a \dmm\ of
1.90.  The CfA spectra have good coverage from five days before
maximum through several weeks past maximum (Figure \ref{99bymont}).  Most of
the spectra were shown and analyzed by \citet{garnavich04}, but the
spectra presented herein were rereduced to be consistent with the rest
of this spectroscopic sample.  This bright and peculiar SN was well
observed by many groups, including \citet{toth00}; \citet{vinko01};
\citet{howell01b}; \citet{hoflich02}; and \citet{garnavich04}.
\clearpage
\clearpage
\emph{SN 1999cc}--\citet{schwartz99} discovered SN~1999cc on 1999 May
8.  Our first CfA spectrum (Figure \ref{99ccmont}), obtained three
days before maximum showed that it was an SN~Ia \citep{garnavich99c}.
Our set of spectra includes a few near maximum and a few three weeks
after maximum.
\clearpage
\clearpage
\emph{SN 1999cl}--This SN was found by the LOSS on 1999 May 29
\citep{papenkova99a}.  Our first CfA spectrum (Figure \ref{99clmont})
indicated that it was of Type Ia, but there was some evidence of
peculiarity \citep{garnavich99d}. The spectrum declined to the blue,
there was a strong \ion{Na}{1}~D absorption at the host galaxy's
velocity (equivalent width [EW] of 3.3 \AA), and there was a
relatively strong Galactic \ion{Na}{1}~D absorption (EW of 0.7 \AA).
These facts suggest that the SN was heavily extinguished by dust.  
In addition, the absorption features of the SN itself were broad, with
the two components of the \ion{S}{2} `W' feature blended together.
Our spectra cover from eight days before to nine days after maximum,
as well as one spectrum about five weeks past maximum.  The \dmm value
of 1.19 \citep{jha06} shows that the faint, reddened nature of this SN
is mainly caused by the intervening extinction.
\clearpage
\clearpage
\emph{SN 1999dq}--Another LOSS discovery, SN~1999dq was found on 1999
Sep 5 \citep{li99a}.  The spectrum was classified as an SN~Ia, but
with some peculiarities \citep{jha99c}.  The \ion{Si}{2} $\lambda$6355
line was shallow and absorptions due to \ion{Fe}{3} were present.
Overall, there was strong similarity to SN~1991T \citep{filippenko92a,
phillips92}.  The spectrum also showed \ion{Na}{1}~D absorption at
the host galaxy's velocity (EW of 1.5 \AA) as well as an \ion{Na}{1}~D
absorption of Galactic origin (EW of 0.8 \AA), suggesting that the SN
suffered from reddening.  The CfA spectra (Figures \ref{99dqmonta} and
\ref{99dqmontb}) start at ten days before maximum and continue almost
daily until a week past maximum.  There are also many spectra one to
three months past maximum.
\clearpage

\begin{figure}
\plotone{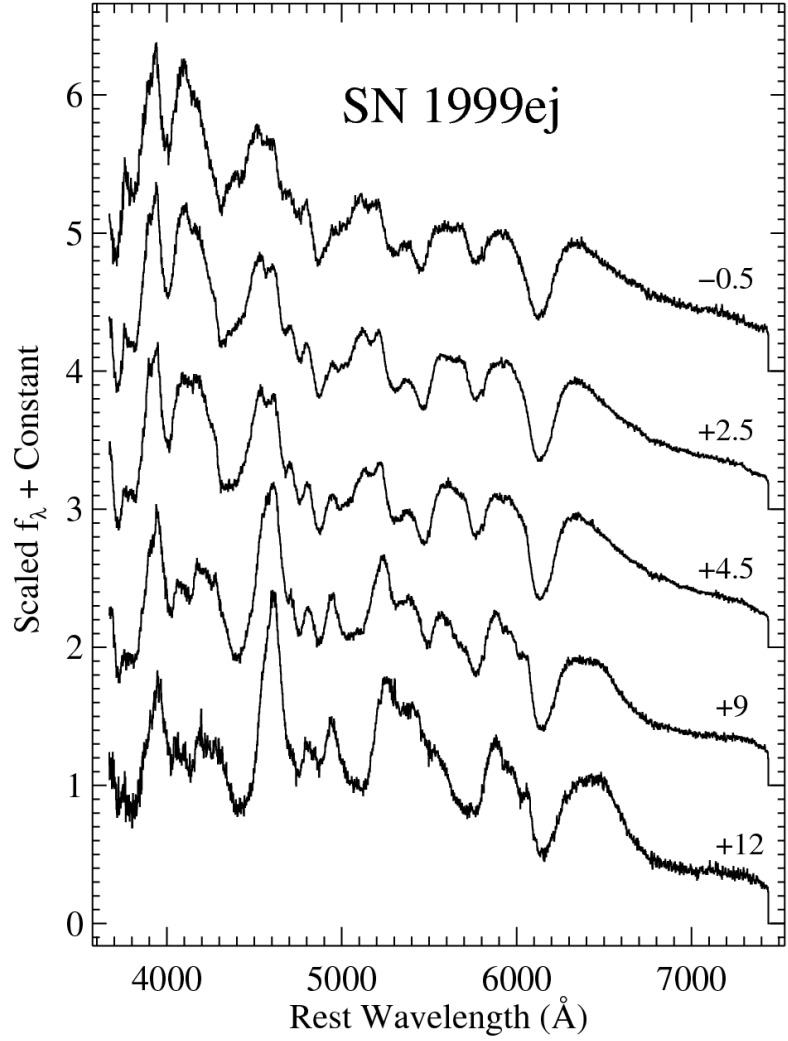}
\caption{Spectra of SN~1999ej.  The flux units, wavelength scale, and
  epoch for each spectrum are as described in Figure
  \ref{97domont}.\label{99ejmont}}
\end{figure}

\begin{figure}
\plotone{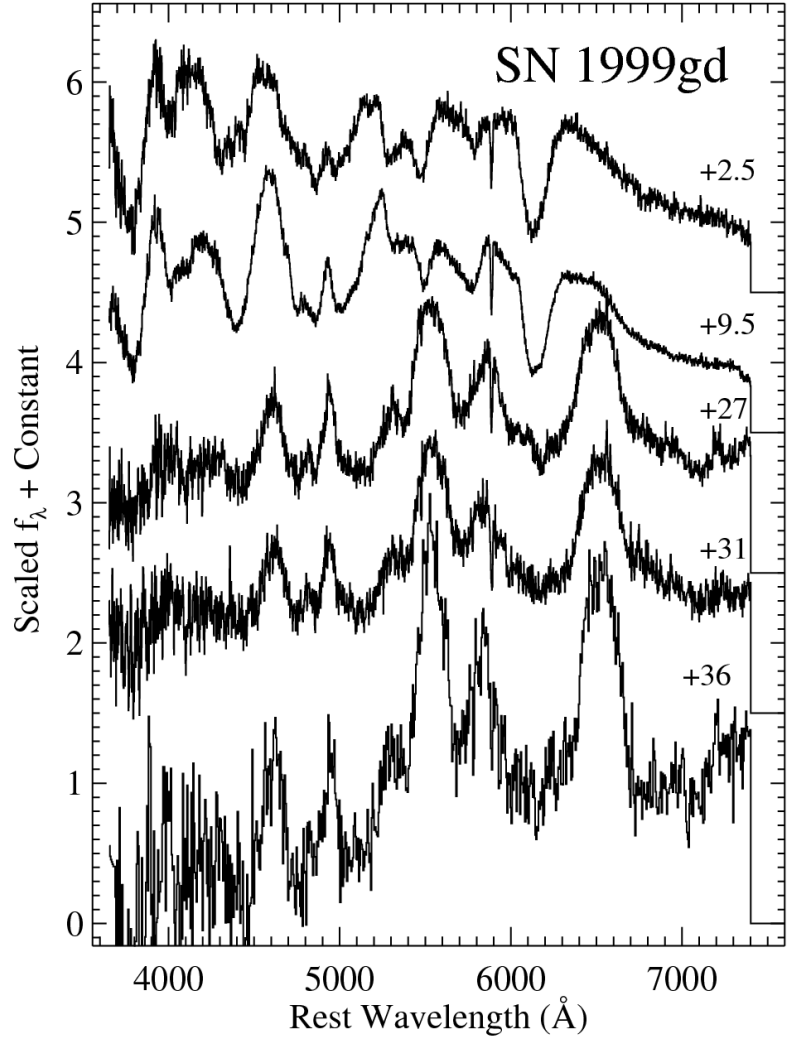}
\caption{Spectra of SN~1999gd.  The flux units, wavelength scale, and
  epoch for each spectrum are as described in Figure \ref{97domont}.
  The day +36 spectrum has been rebinned for clarity.\label{99gdmont}}
\end{figure}

\begin{figure}
\plotone{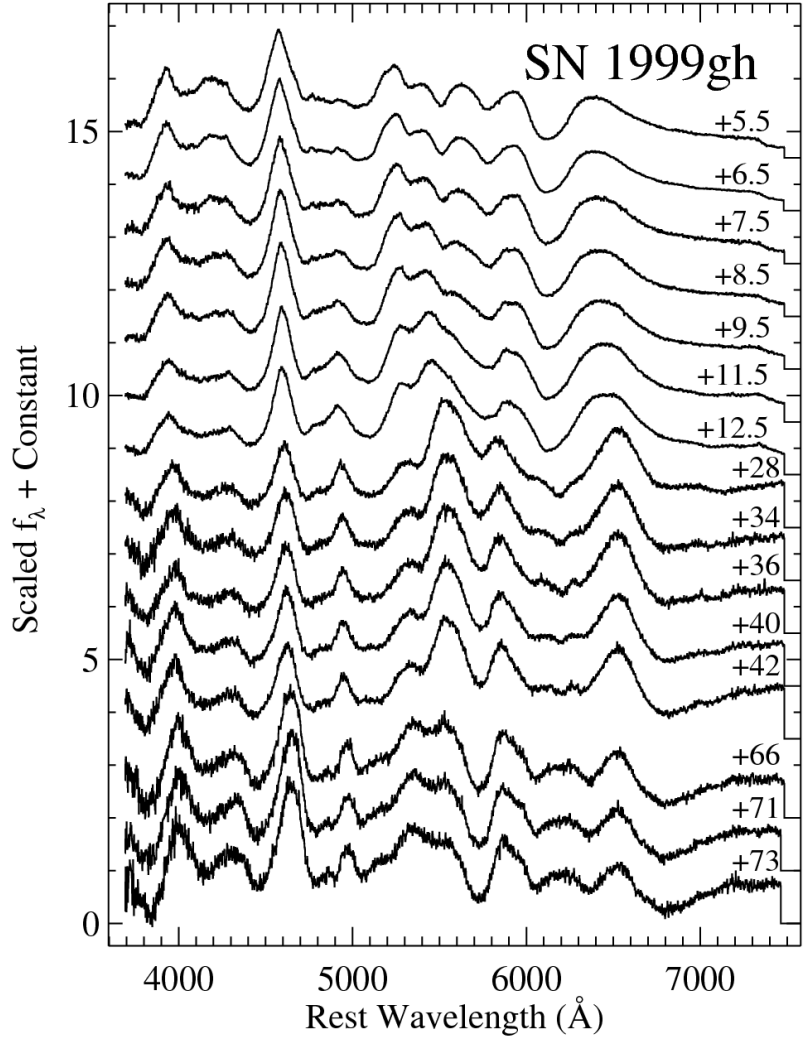}
\caption{Spectra of SN~1999gh.  The flux units, wavelength scale, and
  epoch for each spectrum are as described in Figure
  \ref{97domont}.\label{99ghmont}}
\end{figure}

\begin{figure}
\plotone{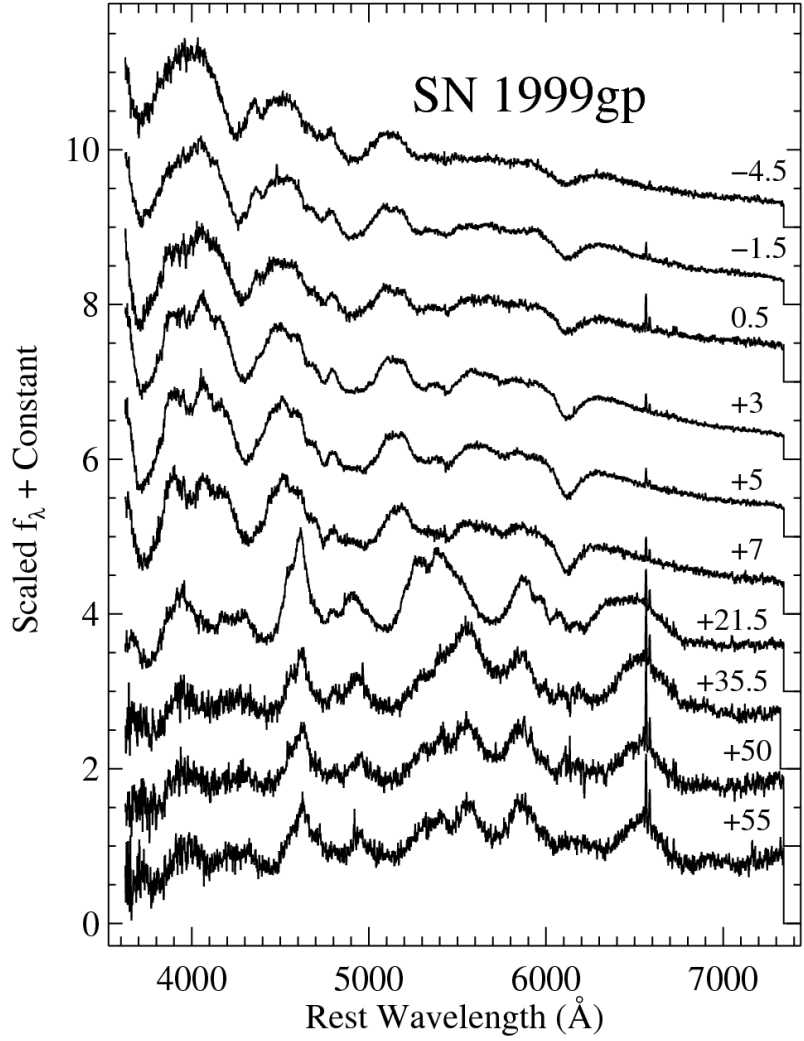}
\caption{Spectra of SN~1999gp.  The flux units, wavelength scale, and
  epoch for each spectrum are as described in Figure
  \ref{97domont}.\label{99gpmont}}
\end{figure}

\begin{figure}
\plotone{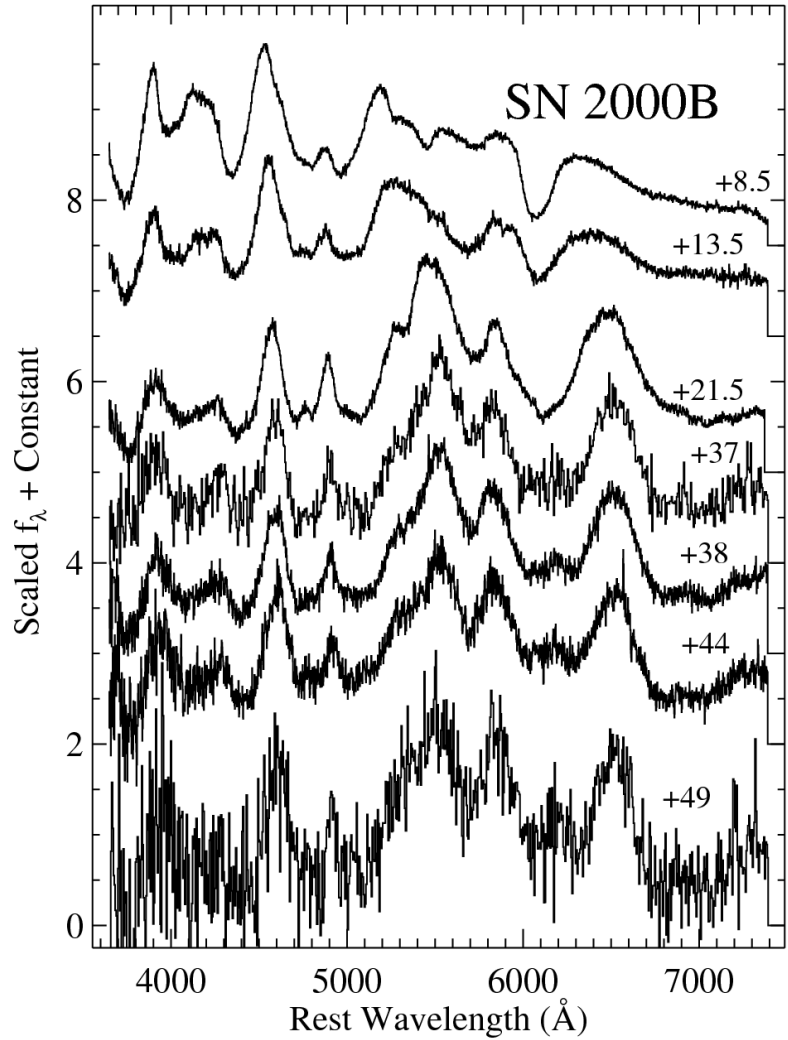}
\caption{Spectra of SN~2000B.  The flux units, wavelength scale, and
  epoch for each spectrum are as described in Figure \ref{97domont}.
  The days +37 and +49 spectra have been rebinned for
  clarity.\label{00bmont}}
\end{figure}

\begin{figure}
\plotone{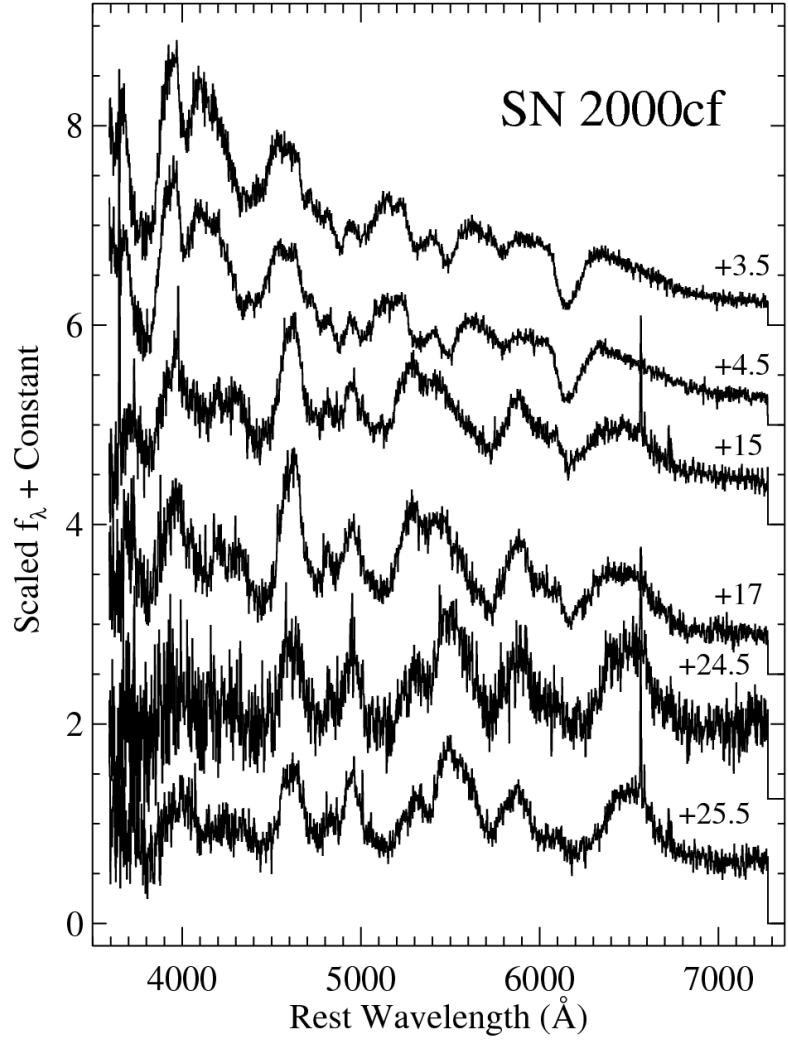}
\caption{Spectra of SN~2000cf.  The flux units, wavelength scale, and
  epoch for each spectrum are as described in Figure
  \ref{97domont}.\label{00cfmont}}
\end{figure}

\begin{figure}
\plotone{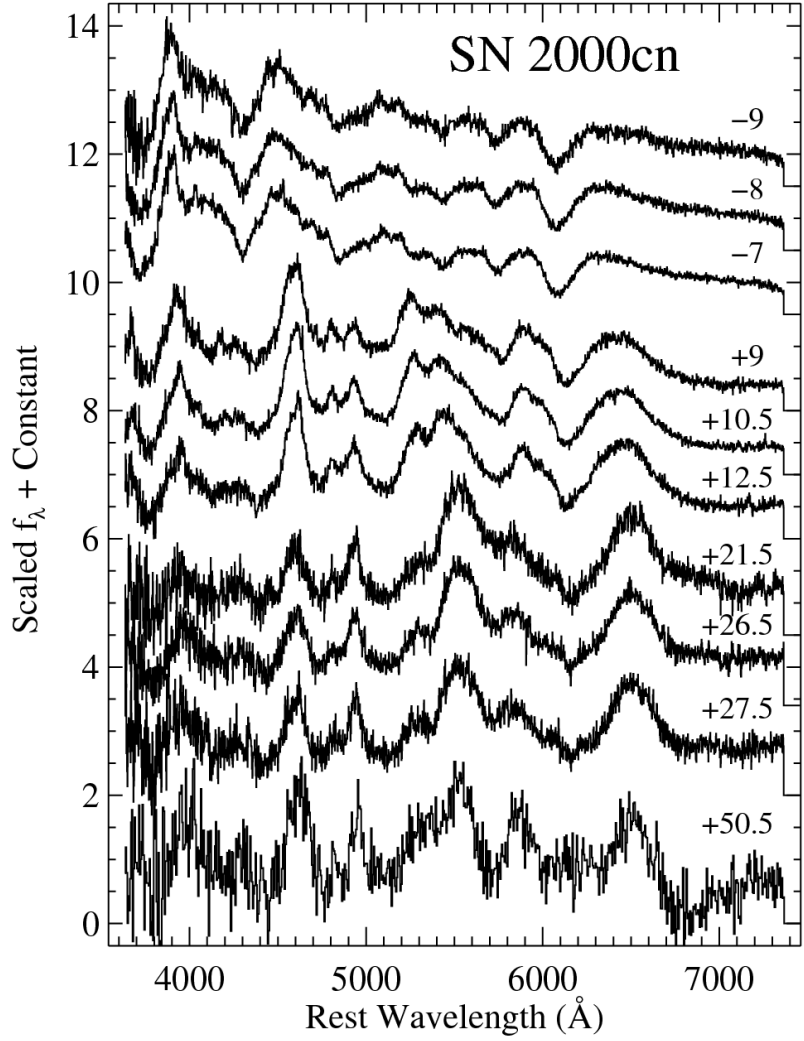}
\caption{Spectra of SN~2000cn.  The flux units, wavelength scale, and
  epoch for each spectrum are as described in Figure
  \ref{97domont}.\label{00cnmont}}
\end{figure}

\begin{figure}
\plotone{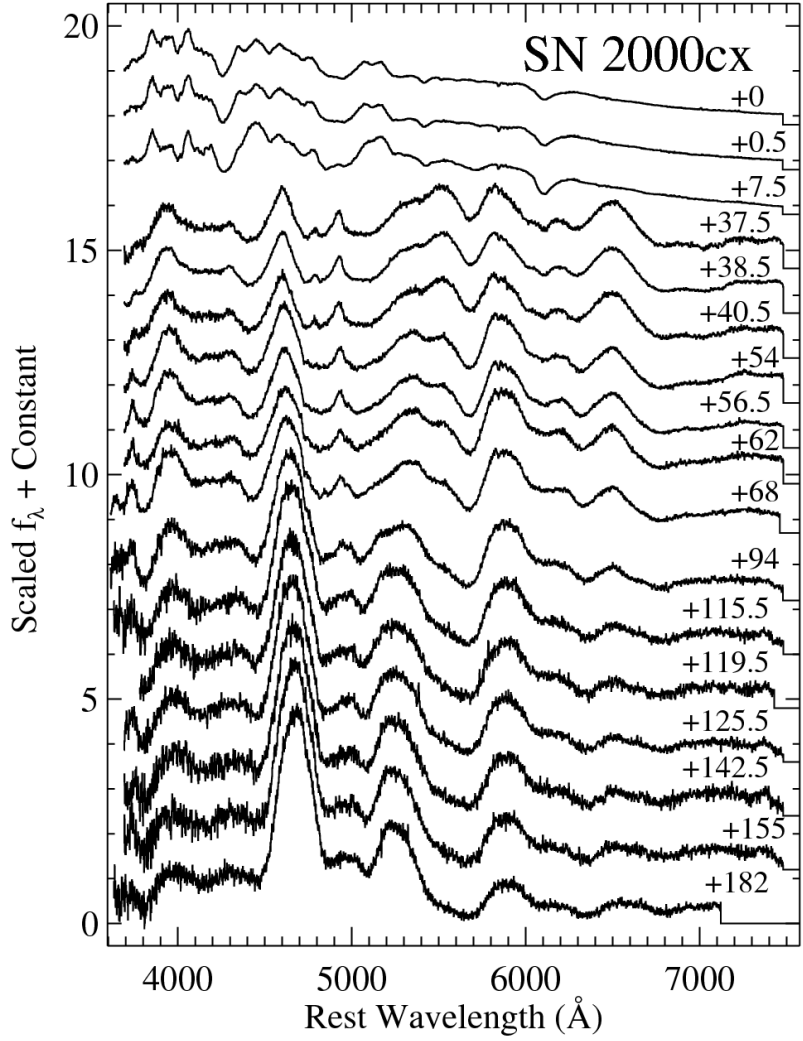}
\caption{Spectra of SN~2000cx.  The flux units, wavelength scale, and
  epoch for each spectrum are as described in Figure
  \ref{97domont}.\label{00cxmont}}
\end{figure}

\begin{figure}
\plotone{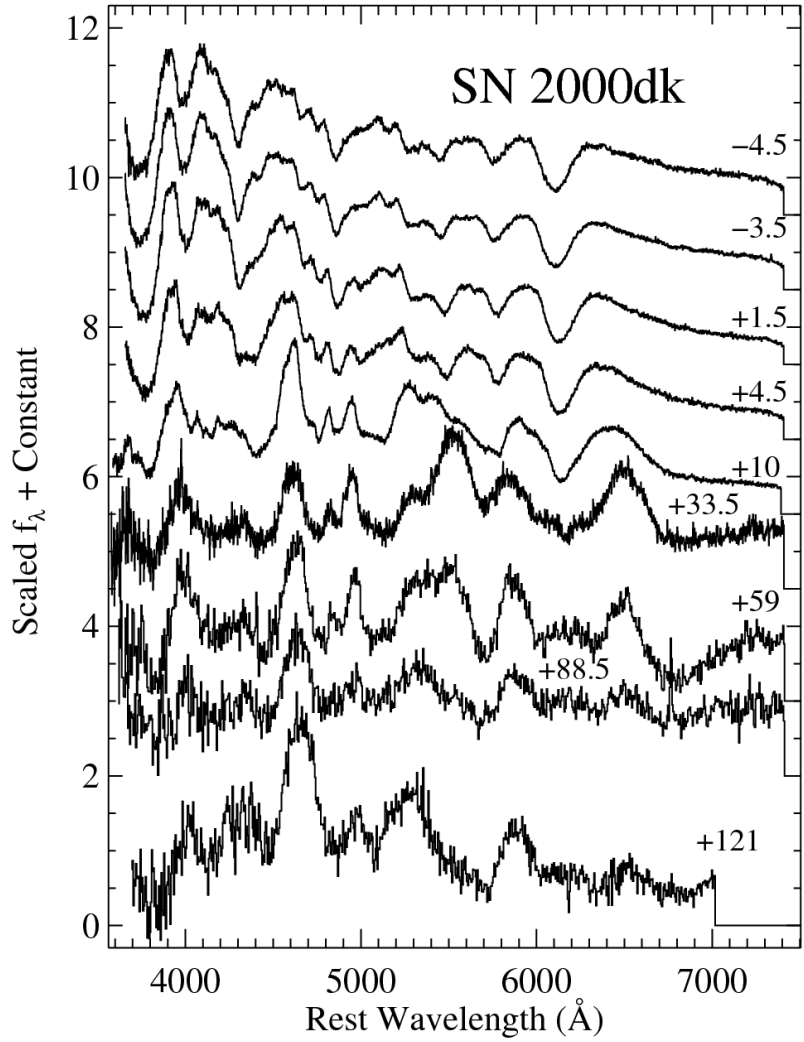}
\caption{Spectra of SN~2000dk.  The flux units, wavelength scale, and
  epoch for each spectrum are as described in Figure \ref{97domont}.
  The days +59, +88.5, and +121 spectra have been rebinned for
  clarity.\label{00dkmont}}
\end{figure}

\begin{figure}
\plotone{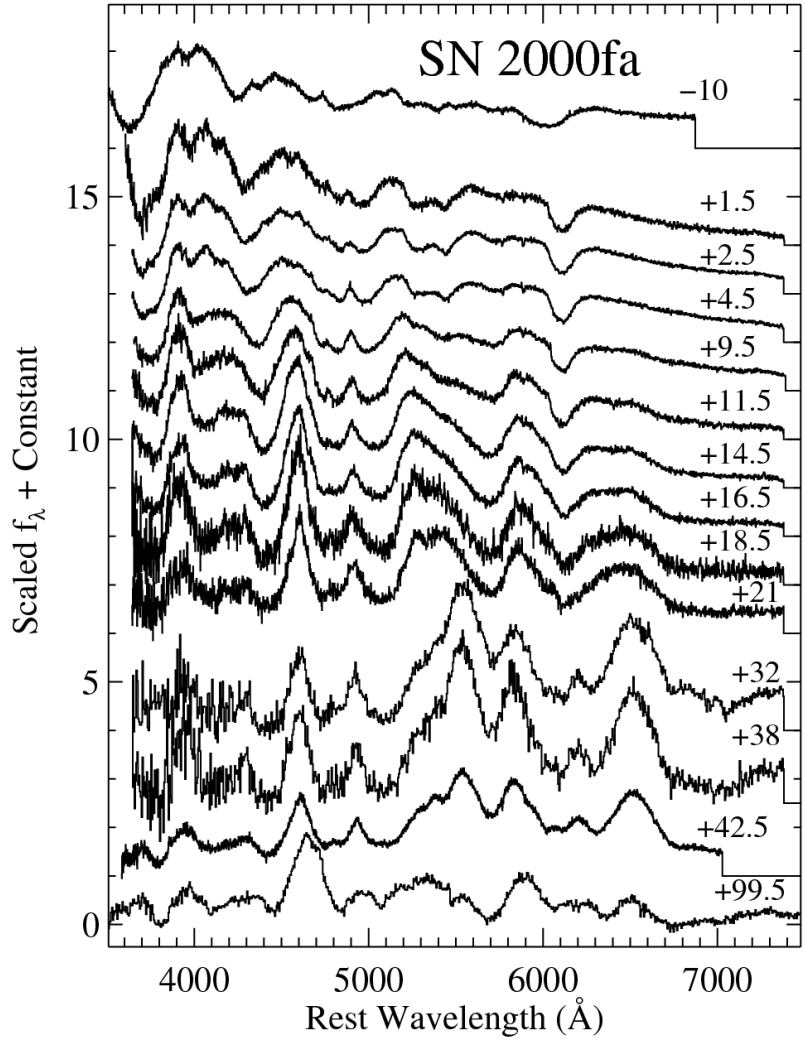}
\caption{Spectra of SN~2000fa.  The flux units, wavelength scale, and
  epoch for each spectrum are as described in Figure \ref{97domont}.
  The days +32, +38, and +99.5 spectra have been rebinned for clarity.
  In addition, the day +99.5 spectrum has been
  rescaled (so the zero point is not indicated).\label{00famont}}
\end{figure}

\begin{figure}
\plotone{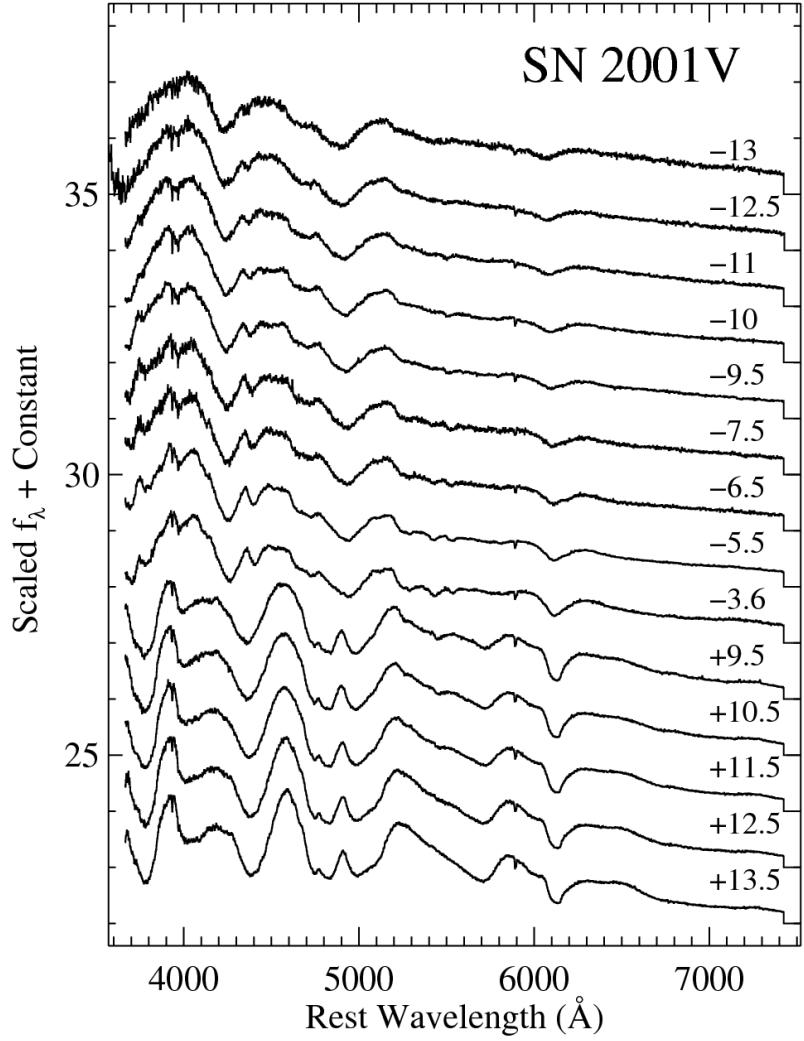}
\caption{Early spectra of SN~2001V.  The flux units, wavelength scale,
  and epoch for each spectrum are as described in Figure
  \ref{97domont}.  \label{01vmonta}}
\end{figure}

\begin{figure}
\plotone{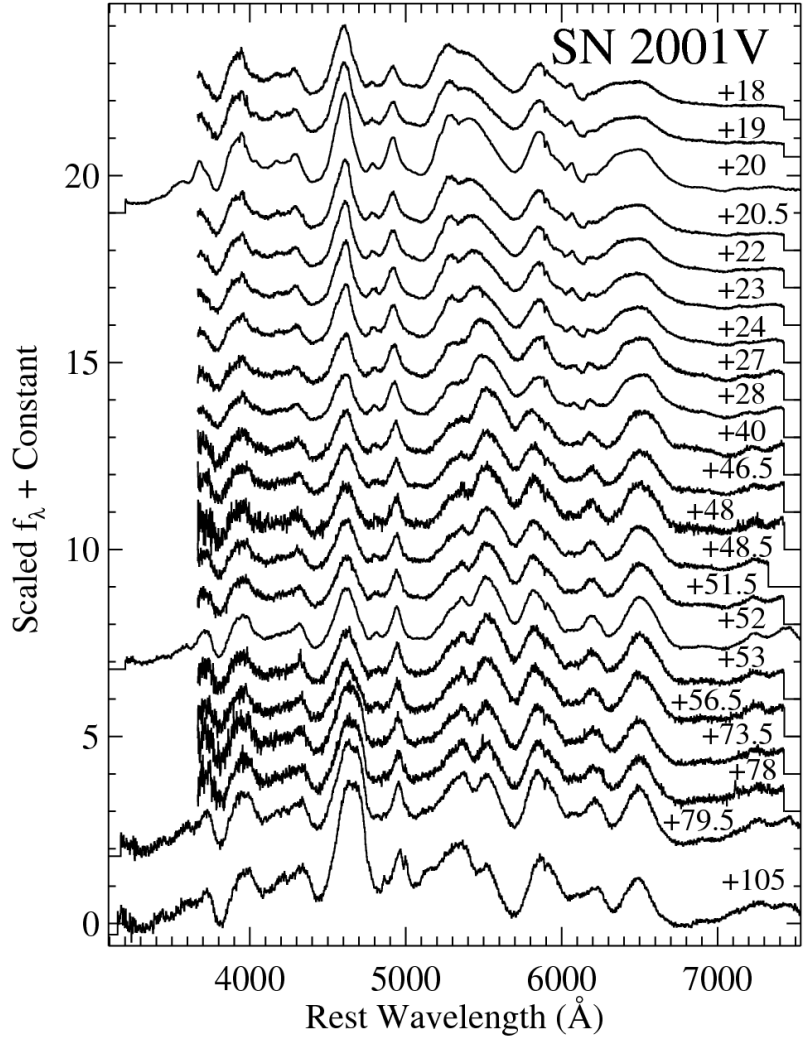}
\caption{Late spectra of SN~2001V.  The flux units, wavelength scale,
  and epoch for each spectrum are as described in Figure
  \ref{97domont}.  \label{01vmontb}}
\end{figure}

\clearpage
\emph{SN 1999ej}--The LOSS also found SN~1999ej on 1999 Oct 18
\citep{friedman99}.  Based upon our first CfA spectrum (Figure
\ref{99ejmont}), obtained one day before maximum, it was classified
as an SN~Ia \citep{jha99a}.  We have a few spectra, from near maximum to
two weeks after maximum.
\clearpage
\clearpage
\emph{SN 1999gd}--The LOSS discovered SN~1999gd on 1999 Nov
24 \citep{li99b}.  A spectrum revealed that it was of Type
Ia and that a strong narrow \ion{Na}{1}~D absorption was present,
  implying that the SN was extinguished by dust \citep{filippenko99b}.  
The \ion{Na}{1}~D line can be seen in the CfA spectra (Figure
  \ref{99gdmont}) and is at the host velocity.  It has an EW of
  4.7\AA.  There are a limited number of spectra, the first at three
  days after maximum, with the rest mainly from a few weeks past maximum.
\clearpage
\clearpage
\emph{SN 1999gh}--\citet{nakano99a} reported the discovery of
SN~1999gh on 1999 Dec 3.  Subsequent spectroscopy showed that it was
an SN ~Ia, and that the \ion{Si}{2} $\lambda$5800 line was stronger
than usual, implying that this might be a subluminous SN~Ia
\citep{filippenko99b}.  We have an extensive set of spectra (Figure
\ref{99ghmont}), but all from well past maximum (starting at six days
after maximum).  The \dmm\ value of 1.69 \citep{jha06} confirms that
this was a peculiar SN~Ia.
\clearpage
\clearpage
\emph{SN 1999gp}--This SN was discovered by the LOSS on 1999 Dec 23
\citep{papenkova99b}.  Our first CfA spectrum (Figure \ref{99gpmont}),
obtained five days before maximum, was used to classify SN~1999gp as
an SN~Ia \citep{jha00a}.  There is good coverage of the post-maximum
decline as well as some spectra several weeks past maximum.
\citet{krisciunas01} present optical and infra-red photometry of this
SN.
\clearpage
\clearpage
\emph{SN 2000B}--\citet{antonini00} found SN~2000B on 2000 Jan 13.
Spectroscopy revealed that it was of Type Ia \citep{colas00}.  The CfA
spectra (Figure \ref{00bmont}) are all from past maximum, starting at
nine days after maximum, with most several weeks past maximum.
\clearpage
\clearpage
\emph{SN 2000cf}--\citet{puckett00} discovered SN2000cf on 2000 May 9.
Our first CfA spectrum (Figure \ref{00cfmont}), obtained four days
after maximum, was used to classify SN~2000cf as an SN~Ia
\citep{jha00c}.  There is limited spectroscopic coverage, all past
maximum.
\clearpage
\clearpage
\emph{SN 2000cn}--The LOSS found SN~2000cn on 2000 Jun 2
\citep{papenkova00}.  Two groups (including the CfA, using our first
spectrum obtained nine days before maximum) reported that
spectroscopy showed that this was an SN~Ia \citep{jha00b}.  Not noted
in those initial reports, but apparent in our spectra (Figure
\ref{00cnmont}), is the relative strength of \ion{Si}{2} $\lambda$5800
line.  Photometry indicated \dmm\ was 1.58 \citep{jha06}, confirming
that this was a subluminous event.  Many of the spectra are from
several weeks past maximum, but there are a few at early epochs.
\clearpage
\clearpage
\emph{SN 2000cx}--Another LOSS discovery, SN~2000cx was found on 2000
Jul 17 \citep{yu00}.  Spectroscopy showed that it was an SN~Ia, but
with some peculiarities \citep{chornock00}.  It resembled SN~1991T
\citep{filippenko92a, phillips92}, with weak \ion{Si}{2} $\lambda$6355
and absorptions due to \ion{Fe}{3}.  This can be seen in the CfA
spectra (Figure \ref{00cxmont}); there are a few near maximum,
starting at maximum, but most are from several weeks to months past
maximum, almost to the onset of the truly nebular phase (again as a
result of the August shutdown at Mt. Hopkins).  As discussed by
\citet{li01a}, SN~2000cx turned out to be so peculiar as to be unique.
The light-curve shape did not match templates, and the spectroscopic
evolution, while similar to SN~1991T, was also distinctive.  This
unusual SN elicited many further analyses \citep{cuadra02, rudy02,
  candia03, branch04, thomas04, sollerman04}.
\clearpage
\clearpage
\emph{SN 2000dk}--The LOSS discovered SN~2000dk on 2000 Sep 18
\citep{beckmann00}.  Our first CfA spectrum (Figure \ref{00dkmont}),
obtained five days before maximum, was used to classify SN~2000dk as
an SN~Ia \citep{jha00d}.  In that report, it was not noted that the
\ion{Si}{2} $\lambda$5800 line was relatively strong and that
\ion{Ti}{2} features were apparent, as can be seen in Figure
\ref{99gpmont}.  The \dmm\ value of 1.57 \citep{jha06}, showed that
this was a subluminous event. We have a few spectra around maximum,
and a few at late times.  \citet{marion03} show an infra-red spectrum
of SN~2000dk.
\clearpage
\clearpage
\emph{SN 2000fa}--Another LOSS discovery, SN~2000fa was found on 2000
Nov 30 \citep{friedman00}.  Using our first CfA spectrum (Figure
\ref{00famont}), this SN was classified as an SN~Ia
 \citep{matheson00b}.  We have many spectra from maximum through
 several weeks past maximum, as well as the first two, both obtained ten days
before maximum.
\clearpage
\clearpage
\emph{SN 2001V}--During the course of a redshift survey program at
Mt. Hopkins, \citet{jha01} discovered SN~2001V spectroscopically on
2001 Feb 19.  This spectrum revealed it to be an SN~Ia, at a very
early epoch, thirteen days before maximum.  We were able to get
extensive spectroscopic coverage of this SN (Figures \ref{01vmonta}
and \ref{01vmontb}) from 13 days before maximum to more than three
months past maximum.  \citet{vinko03} also present photometry of
SN~2001V.
\clearpage

\clearpage

\section{Conclusions}

We have presented a large, homogeneous set of low-dispersion optical
spectra of SNe~Ia.  All the SNe have well-calibrated light curves with
known properties such as the decline rate.  The consistency of
observation and reduction makes this an ideal sample for studying
spectroscopic characteristics of SNe~Ia in relation to the nature of
their light curves (e.g., Matheson et al. 2008, in preparation).

\acknowledgments

We would like to thank the staffs of the F.~L. Whipple and MMT
Observatories for their extensive assistance and support during this
project.  We would also like to thank Dan Koranyi and Barbara Carter
for providing assistance with the observations.  The CfA time
allocation committee has been extremely generous in supporting SN
observations at FLWO, especially in allowing the day-by-day monitoring
of so many SNe.  This research was supported in part by the National
Science Foundation under Grant No. PHY05-51164 to the KITP and
AST-0606772 to Harvard University.

\clearpage
\begin{deluxetable}{llrrc}
\tablecaption{SN Ia and Host Basic Data\label{sndata}}
\tablehead{\colhead{SN} &
\colhead{Host} &
\colhead{c$z_{\rm{helio}}$\tablenotemark{a}} &
\colhead{Host morph.\tablenotemark{b}} &
\colhead{$E(B-V)_{\rm{Galactic}}$\tablenotemark{c}} \\
\colhead{} &
\colhead{ galaxy} &
\colhead{km~s$^{-1}$} &
\colhead{} &
\colhead{mag} }
\startdata
SN~1997do  &UGC 3845       &3034    &Sbc    &0.063    \\
SN~1997dt  &NGC 7448       &2194    &Sbc    &0.057   \\
SN~1998V   &NGC 6627       &5268    &Sb     &0.196   \\
SN~1998ab  &NGC 4704       &8134    &Sc     &0.017   \\
SN~1998aq  &NGC 3982	   &1184    &Sb	    &0.014 \\
SN~1998bp  &NGC 6495       &3127    &E      &0.076   \\
SN~1998bu  &NGC 3368	   &897	    &Sab    &0.025 \\
SN~1998de  &NGC 252        &4990    &S0     &0.057   \\
\enddata
\tablenotetext{a}{Heliocentric redshifts listed are from the Updated
  Zwicky Catalog \citep{falco99} if possible, and from the NASA/IPAC
  Extragalactic Database (NED) if not.  For the host of SN~2000cf
  (MCG+11-19-25), we derived the velocity of 10920 km~s$^{-1}$ from
  our own spectra of the host galaxy.}
\tablenotetext{b}{Host galaxy morphology is taken from NED, if
  possible, and from the LEDA database if not.}
\tablenotetext{c}{The Galactic reddening toward each SN is derived
  from the dust maps of \citet{schlegel98}.}
\end{deluxetable}

\begin{deluxetable}{lcccccccccccccl}
\tabletypesize{\scriptsize}
\rotate
%\tablenum{1}
\tablewidth{0pt}
\tablecaption{JOURNAL OF OBSERVATIONS\label{obslog}}
\tablehead{\colhead{} & 
\colhead{UT Date\tablenotemark{a}} &
\colhead{HJD\tablenotemark{b}} &
\colhead{Phase\tablenotemark{c}} &
\colhead{Tel.\tablenotemark{d}} &
\colhead{Range}  &
\colhead{Res.\tablenotemark{e}} &
\colhead{P.A.\tablenotemark{f}} &
\colhead{Par.\tablenotemark{g}} &
\colhead{Air.} & 
\colhead{Flux Std.\tablenotemark{h}} &
\colhead{See.\tablenotemark{i}} &
\colhead{Slit} &
\colhead{Exp.} &
\colhead{Observer(s)\tablenotemark{j}} \\
\colhead{} &
\colhead{} &
\colhead{} &
\colhead{} &
\colhead{} &
\colhead{(\AA)} &
\colhead{(\AA)} &
\colhead{($^\circ$)} &
\colhead{($^\circ$)} &
\colhead{} &
\colhead{} &
\colhead{($^{\prime\prime}$)} &
\colhead{($^{\prime\prime}$)} &
\colhead{(s)} & 
\colhead{} }
\startdata
SN 1997do & & & & & & & & & & & & & \\
 & 1997-11-02.45 & 2450754.95 & -11 & FLWO & 3720-7540.5 & 7.0 & 90.0 & -128.7 & 1.07 & F34 & 1-2 & 3 & 600 & DK \\
 & 1997-11-03.42 & 2450755.92 & -10 & FLWO & 3720-7540.5 & 7.0 & 90.0 & -114.7 & 1.11 & F34 & 1 & 3 & 600 & DK \\
 & 1997-11-06.49 & 2450758.99 & -7 & FLWO & 3720-7540.5 & 7.0 & 90.0 & -177.4 & 1.04 & F34 & 1-2 & 3 & 1200 & PB \\
 & 1997-11-07.49 & 2450760.00 & -6 & FLWO & 3720-7540.5 & 7.0 & 90.0 & 172.3 & 1.04 & F34 & 1-2 & 3 & 900 & PB \\
 & 1997-11-22.44 & 2450774.95 & 8.5 & FLWO & 3720-7540.5 & 7.0 & 90.0 & -172.3 & 1.04 & F34 & 1-2 & 2 & 1200+900 & PB, JK, BW \\
 & 1997-11-24.44 & 2450776.94 & 10.5 & FLWO & 3720-7540.5 & 7.0 & 90.0 & -173.3 & 1.04 & F34 & 1-2 & 3 & 900 & PB \\
 & 1997-11-25.36 & 2450777.87 & 11.5 & FLWO & 3720-7540.5 & 7.0 & 90.0 & -116.2 & 1.09 & F34 & 2 & 3 & 1020 & PB \\
\enddata
\end{deluxetable}

\clearpage


\begin{thebibliography}{}

\bibitem[Antonini et al.(2000)]{antonini00} Antonini, P., Colas, F.,
Frappa, E., \& Li, W.\ 2000, \iaucirc\, 7347

\bibitem[Arbour \& Armstrong(1999)]{arbour99a} Arbour, R., \&
Armstrong, M. 1999, \iaucirc\, 7108

\bibitem[Arbour et al.(1999)]{arbour99b} Arbour, R., Papenkova, M., Li,
W.~D., Filippenko, A.~V., \& Armstrong, M.\ 1999, \iaucirc\, 7156

\bibitem[Astier et al.(2006)]{astier06} Astier, P., et al.\ 
2006, \aap, 447, 31 

\bibitem[Ayani et al.(1998)]{ayani98a} Ayani, K., et al.\ 1998,
\iaucirc\, 6905

\bibitem[Ayani \& Yamaoka(1998)]{ayani98b} Ayani, K., \& Yamaoka,
H. 1998, \iaucirc\, 6878

\bibitem[Beckmann \& Li(2000)]{beckmann00} Beckmann, S., \& Li, W.~D.\
2000, \iaucirc\, 7493

\bibitem[Benetti et al.(2004)]{benetti04} Benetti, S., et al.\ 
2004, \mnras, 348, 261

\bibitem[Benetti et al.(2005)]{benetti05} Benetti, S., et al.\ 
2005, \apj, 623, 1011 
 

\bibitem[Bessell(1999)]{bessell99} Bessell, M.~S.\ 1999, \pasp, 111,
1426

\bibitem[Branch(2001)]{branch01} Branch, D.\ 2001, \pasp, 113, 
169

\bibitem[Branch et al.(2003)]{branch03} Branch, D., et al.\ 2003, \aj,
126, 1489

\bibitem[Branch et al.(2004)]{branch04} Branch, D., et al.\ 2004,
\apj, 606, 413

\bibitem[Branch et al.(2005)]{branch05} Branch, D., et al.\ 2005,
\pasp, 117, 545

\bibitem[Candia et al.(2003)]{candia03} Candia, P., et al.\ 2003,
\pasp, 115, 277

\bibitem[Chornock et al.(2000)]{chornock00} Chornock, R., Leonard,
D.~C., Filippenko, A.~V., Li, W.~D., Gates, E.~L., \& Chloros, K.\
2000, \iaucirc, 7463

\bibitem[Colas et al.(2000)]{colas00} Colas, F., Fienga, A., \& Buil,
C.\ 2000, \iaucirc\, 7351

\bibitem[Cuadra et al.(2002)]{cuadra02} Cuadra, J., Suntzeff, N.~B.,
Candia, P., Krisciunas, K., \& Phillips, M.~M.\ 2002, Revista Mexicana
de Astronomia y Astrofisica Conference Series, 14, 121

\bibitem[Fabricant et al.(1998)]{fabricant98} Fabricant, D., Cheimets,
P., Caldwell, N., \& Geary, J.\ 1998, \pasp, 110, 79

\bibitem[Falco et al.(1999)]{falco99} Falco, E.~E., et al. 1999,
\pasp, 111, 438

\bibitem[Filippenko(1982)]{filippenko82} Filippenko, A. V. 1982,
\pasp, 94, 715

%\bibitem[Filippenko(1997)]{filippenko97} Filippenko, A.~V. 1997,
%\araa, 35, 309

\bibitem[Filippenko \& De Breuck(1998)]{filippenko98} Filippenko,
A.~V., \& De Breuck, C.\ 1998, \iaucirc\, 6997

\bibitem[Filippenko et al.(1992a)]{filippenko92a} Filippenko, A.~V., 
et al.\ 1992a, \apjl, 384, L15

\bibitem[Filippenko et al.(1992b)]{filippenko92b} Filippenko, A.~V., et
al.\ 1992b, \aj, 104, 1543

\bibitem[Filippenko et al.(1999)]{filippenko99a} Filippenko, A.~V., Li,
W.~D., \& Leonard, D.~C. 1999, \iaucirc\, 7108

\bibitem[Filippenko \& Garnavich(1999)]{filippenko99b} Filippenko,
A.~V., \& Garnavich, P.\ 1999, \iaucirc\, 7328 %both 99gh and 99gd

\bibitem[Friedman et al.(1999)]{friedman99} Friedman, A., King, J.~Y.,
\& Li, W.~D.\ 1999, \iaucirc\, 7286

\bibitem[Friedman et al.(2000)]{friedman00} Friedman, A., Li, W.~D.,
\& Schwartz, M.\ 2000, \iaucirc\, 7533

\bibitem[Gamezo et al.(2004)]{gamezo04} Gamezo, V.~N., Khokhlov, 
A.~M., \& Oran, E.~S.\ 2004, Physical Review Letters, 92, 211102

\bibitem[Garavini et al.(2004)]{garavini04} Garavini, G., et al.\
2004, \aj, 128, 387

\bibitem[Garnavich et al.(1998a)]{garnavich98a} Garnavich, P., Jha, S.,
Kirshner, R. \& Berlind, P.  1998a, \iaucirc\, 6980

\bibitem[Garnavich et al.(1998b)]{garnavich98b} Garnavich, P., Jha, S.,
Kirshner, R., Berlind, P., \& Calkins, M. 1998b, \iaucirc\, 6858

\bibitem[Garnavich et al.(1998c)]{garnavich98c} Garnavich, P., Jha, S.,
Kirshner, R., Berlind, P., \& Calkins, M. 1998c, \iaucirc\, 6880

\bibitem[Garnavich et al.(1998d)]{garnavich98d} Garnavich, P., Jha, S.,
Kirshner, R., \& Calkins, M.\ 1998d, \iaucirc\, 6980

\bibitem[Garnavich et al.(1999a)]{garnavich99a} Garnavich, P., Jha, S.,
Kirshner, R. \& Berlind, P. 1999a, \iaucirc\, 7105

\bibitem[Garnavich et al.(1999b)]{garnavich99b} Garnavich, P., Jha, S.,
Kirshner, R., \& Berlind, P.\ 1999b, \iaucirc\, 7159

\bibitem[Garnavich et al.(1999c)]{garnavich99c} Garnavich, P., Jha, S.,
Kirshner, R., Challis, P., \& Berlind, P.\ 1999c, \iaucirc\, 7169

\bibitem[Garnavich et al.(1999d)]{garnavich99d} Garnavich, P., Jha, S.,
Kirshner, R., Challis, P., \& Calkins, M.\ 1999d, \iaucirc\, 7190

\bibitem[Garnavich et al.(2004)]{garnavich04} Garnavich, P.~M., et
al.\ 2004, \apj, 613, 1120

\bibitem[Gerardy \& Fesen(1999)]{gerardy99} Gerardy, C., \& Fesen, R.\
1999, \iaucirc\, 7158

%\bibitem[Goldhaber et al.(2001)]{goldhaber01} Goldhaber, G., et
%  al. 2001, \apj, 558, 359

\bibitem[Halderson et al.(1998)]{halderson98} Halderson, E., Modjaz,
M., Shefler, T., King, J.~Y., Papenkova, M., Li, W.~D., Treffers,
R.~R., \& Filippenko, A.~V.\ 1998, \iaucirc\, 7050

\bibitem[Hamuy et al.(1996a)]{hamuy96a} Hamuy, M., Phillips, 
M.~M., Suntzeff, N.~B., Schommer, R.~A., Maza, J., \& Aviles, R.\ 1996a,
\aj, 112, 2391 

\bibitem[Hamuy et al.(1996b)]{hamuy96b} Hamuy, M., et al.\ 1996b, 
\aj, 112, 2408 

\bibitem[Hatano et al.(2000)]{hatano00} Hatano, K., Branch, D., 
Lentz, E.~J., Baron, E., Filippenko, A.~V., \& Garnavich, P.~M.\ 2000, 
\apjl, 543, L49 

\bibitem[Hillebrandt \& Niemeyer(2000)]{hillebrandt00} Hillebrandt, 
W., \& Niemeyer, J.~C.\ 2000, \araa, 38, 191

\bibitem[H{\" o}flich et al.(2002)]{hoflich02} H{\" o}flich, P.,
Gerardy, C.~L., Fesen, R.~A., \& Sakai, S.\ 2002, \apj, 568, 791

\bibitem[Horne(1986)]{horne86}Horne, K. 1986, \pasp, 98, 609

\bibitem[Howell(2001a)]{howell01a} Howell, D.~A.\ 2001a, \apjl, 
554, L193

\bibitem[Howell et al.(2001b)]{howell01b} Howell, D.~A., H{\" o}flich,
P., Wang, L., \& Wheeler, J.~C.\ 2001b, \apj, 556, 302

\bibitem[Hurst \& Armstrong(1998)]{hurst98d} Hurst, G.~M., \&
Armstrong, M. 1998, \iaucirc\, 6890

\bibitem[Hurst et al.(1998a)]{hurst98a} Hurst, G.~M., Armstrong, M., \&
Arbour, R. 1998a, \iaucirc\, 6875

\bibitem[Hurst et al.(1998b)]{hurst98b} Hurst, G.~M., Armstrong, M.,
Boles, T., Nakano, S., Kushida, Y., \& Kushida, R.\ 1998b, \iaucirc\,
6841

\bibitem[Hurst et al.(1998c)]{hurst98c} Hurst, G.~M., et al.\ 1998c,
\iaucirc, 7033

\bibitem[Jha et al.(1998a)]{jha98a} Jha, S., Garnavich, P., Challis, P.,
Kirshner, R., \& Berlind, P.\ 1998a, \iaucirc\, 7024

\bibitem[Jha et al.(1998b)]{jha98b} Jha, S., Garnavich, P., Challis, P.,
Kirshner, R., \& Calkins, M. 1998b, \iaucirc\, 6891

\bibitem[Jha et al.(1998c)]{jha98c} Jha, S., Garnavich, P., Challis, P.,
Kirshner, R., \& Calkins, M.\ 1998c, \iaucirc\, 7054

\bibitem[Jha et al.(1998d)]{jha98d} Jha, S., Garnavich, P., Kirshner,
R., Koranyi, D., \& Calkins, M.\ 1998d, \iaucirc\, 6844

\bibitem[Jha et al.(1999a)]{jha99a} Jha, S., Challis, P., Garnavich, P.,
Kirshner, R., \& Calkins, M.\ 1999a, \iaucirc\, 7298

\bibitem[Jha et al.(1999b)]{jha99b} Jha, S., et al.\ 1999b, \apjs, 125,
73

\bibitem[Jha et al.(1999c)]{jha99c} Jha, S., Garnavich, P., Challis, P.,
Kirshner, R., \& Berlind, P.\ 1999c, \iaucirc\, 7250

\bibitem[Jha et al.(2000a)]{jha00a} Jha, S., Challis, P., Kirshner, R.,
\& Berlind, P.\ 2000a, \iaucirc\, 7341

\bibitem[Jha et al.(2000b)]{jha00b} Jha, S., Challis, P., Kirshner, R.,
Berlind, P., Turatto, M., Pastorello, A., Cappellaro, E., \& Cedrati,
F.\ 2000b, \iaucirc\, 7437

\bibitem[Jha et al.(2000c)]{jha00c} Jha, S., Challis, P., Kirshner, R.,
\& Calkins, M.\ 2000c, \iaucirc\, 7423

\bibitem[Jha et al.(2000d)]{jha00d} Jha, S., Challis, P., Matheson, T.,
Kirshner, R., \& Berlind, P.\ 2000d, \iaucirc\, 7494

\bibitem[Jha et al.(2001)]{jha01} Jha, S., Matheson, T., Challis, P., 
Kirshner, R., \& Berlind, P.\ 2001, \iaucirc\, 7585

\bibitem[Jha et al.(2007)]{jha07} Jha, S., Riess, A.~G., \& 
Kirshner, R.~P.\ 2007, \apj, 659, 122 

\bibitem[Jha et al.(2006)]{jha06} Jha, S., et al.\ 2006, \aj, 
131, 527 

\bibitem[King et al.(1998)]{king98} King, J.~Y., Modjaz, M., Shefler,
T., Halderson, E., Li, W.~D., Treffers, R.~R., \& Filippenko, A.~V.\
1998, \iaucirc\, 6991

\bibitem[Kirshner et al.(1973)]{kirshner73} Kirshner, R.~P., Oke, 
J.~B., Penston, M.~V., \& Searle, L.\ 1973, \apj, 185, 303 

\bibitem[Kirshner \& Oke(1975)]{kirshner75} Kirshner, R.~P., \& 
Oke, J.~B.\ 1975, \apj, 200, 574 

\bibitem[Kirshner et al.(1993)]{kirshner93} Kirshner, R.~P., et 
al.\ 1993, \apj, 415, 589

\bibitem[Knop et al.(2003)]{knop03} Knop, R.~A., et al.\ 2003, 
\apj, 598, 102

\bibitem[Kotak et al.(2005)]{kotak05} Kotak, R., et al.\ 2005, 
\aap, 436, 1021

\bibitem[Krisciunas et al.(2001)]{krisciunas01} Krisciunas, K., et
al.\ 2001, \aj, 122, 1616

\bibitem[Krisciunas et al.(2003)]{krisciunas03} Krisciunas, K., et 
al.\ 2003, \aj, 125, 166

\bibitem[Kuchner et al.(1994)]{kuchner94} Kuchner, M.~J., 
Kirshner, R.~P., Pinto, P.~A., \& Leibundgut, B.\ 1994, \apjl, 426, L89

\bibitem[Leibundgut et al.(1993)]{leibundgut93} Leibundgut, B., et
al.\ 1993, \aj, 105, 301

\bibitem[Li(1999a)]{li99a} Li, W.~D.\ 1999a, \iaucirc\, 7247

\bibitem[Li(1999b)]{li99b} Li, W.~D.\ 1999b, \iaucirc\, 7319

\bibitem[Li et al.(1998)]{li98} Li, W.~D., Modjaz, M., Halderson, E.,
Shefler, T., King, J.~Y., Treffers, R.~R., \& Filippenko, A.~V.\ 1998,
\iaucirc\, 6978

\bibitem[Li et al.(2001a)]{li01a} Li, W., et al.\ 2001a, \pasp, 113, 1178

\bibitem[Li et al.(2001b)]{li01b} Li, W., Filippenko, A.~V., Treffers,
R.~R., Riess, A.~G., Hu, J., \& Qiu, Y.\ 2001b, \apj, 546, 734

\bibitem[Mandel et al.(2008)]{mandel07} Mandel, K., et al. 2008, in
  preparation 

\bibitem[Marion et al.(2003)]{marion03} Marion, G.~H., H{\" o}flich,
P., Vacca, W.~D., \& Wheeler, J.~C.\ 2003, \apj, 591, 316

\bibitem[Massel et al.(1988)]{massey88} Massey, P., et al. 1988, \apj,
  328, 315

\bibitem[Massey \& Gronwall(1990)]{massey90} Massey, P., \& Gronwall,
  C. 1990, \apj, 358, 344

\bibitem[Matheson et al.(2000a)]{matheson00a} Matheson, T., Filippenko,
A.~V., Ho, L.~C., Barth, A.~J., \& Leonard, D.~C.\ 2000a, \aj, 120,
1499

\bibitem[Matheson et al.(2000b)]{matheson00b} Matheson, T., Jha, S.,
Challis, P., Kirshner, R., Huchra, J., \& Caldwell, N.\ 2000b,
\iaucirc\, 7535

\bibitem[Matheson et al.(2008)]{matheson07} Matheson, T., et
  al. 2008, in preparation

\bibitem[Mazzali et al.(1998)]{mazzali98} Mazzali, P.~A., 
Cappellaro, E., Danziger, I.~J., Turatto, M., \& Benetti, S.\ 1998, \apjl, 
499, L49 

\bibitem[Modjaz et al.(1998a)]{modjaz98a} Modjaz, M., Halderson, E.,
Shefler, T., King, J.~Y., Li, W.~D., Treffers, R.~R., \& Filippenko,
A.~V.\ 1998a, \iaucirc\, 6977

\bibitem[Modjaz et al.(1998b)]{modjaz98b} Modjaz, M., Shefler, T.,
Halderson, E., King, J.~Y., Li, W.~D., Treffers, R.~R., \& Filippenko,
A.~V.\ 1998b, \iaucirc\, 6993

\bibitem[Modjaz et al.(1999)]{modjaz99} Modjaz, M., King, J.~Y.,
Papenkova, M., Friedman, A., Johnson, R.~A., Li, W.~D., Treffers,
R.~R., \& Filippenko, A.~V.\ 1999, \iaucirc\, 7114

\bibitem[Modjaz et al.(2001)]{modjaz01} Modjaz, M., Li, W.,
Filippenko, A.~V., King, J.~Y., Leonard, D.~C., Matheson, T.,
Treffers, R.~R., \& Riess, A.~G.\ 2001, \pasp, 113, 308

\bibitem[Nakano et al.(1999)]{nakano99a} Nakano, S., Takamizawa, K.,
Kushida, Y., \& Kushida, R.\ 1999, \iaucirc, 7328

\bibitem[Nakano \& Kushida(1999)]{nakano99b} Nakano, S., \& Kushida,
R. 1999, \iaucirc\, 7109

\bibitem[Nomoto et al.(2003)]{nomoto03} Nomoto, K., Uenishi, T., 
Kobayashi, C., Umeda, H., Ohkubo, T., Hachisu, I., \& Kato, M.\ 2003, From 
Twilight to Highlight: The Physics of Supernovae, 115 

\bibitem[Oke(1974)]{oke74} Oke, J.~B. 1974, \apjs, 27, 21

\bibitem[Oke(1990)]{oke90} Oke, J.~B.\ 1990, \aj, 99, 1621 

\bibitem[Oke \& Gunn(1983)]{oke83} Oke, J.~B., \& Gunn, J.~E. 1983,
  \apj, 266, 713

\bibitem[Papenkova et al.(1999)]{papenkova99a} Papenkova, M.,
Filippenko, A.~V., \& Treffers, R.~R.\ 1999a, \iaucirc\, 7185

\bibitem[Papenkova \& Li(1999)]{papenkova99b} Papenkova, M., \& Li,
W.~D.\ 1999b, \iaucirc\, 7337

\bibitem[Papenkova \& Li(2000)]{papenkova00} Papenkova, M., \& Li,
W.~D.\ 2000, \iaucirc\, 7436

\bibitem[Patat \& Maia(1998)]{patat98} Patat, F., \& Maia, M. 1998,
\iaucirc\, 6890

\bibitem[Perlmutter et al.(1997)]{perlmutter97} Perlmutter, S., et 
al.\ 1997, \apj, 483, 565

\bibitem[Perlmutter et al.(1999)]{perlmutter99} Perlmutter, S., et 
al.\ 1999, \apj, 517, 565

\bibitem[Phillips(1993)]{phillips93} Phillips, M.~M.\ 1993, \apjl, 
413, L105

\bibitem[Phillips et al.(1992)]{phillips92} Phillips, M.~M., Wells,
L.~A., Suntzeff, N.~B., Hamuy, M., Leibundgut, B., Kirshner, R.~P., \&
Foltz, C.~B.\ 1992, \aj, 103, 1632

\bibitem[Phillips et al.(1999)]{phillips99} Phillips, M.~M., Kunkel,
W., \& Filippenko, A.~V.\ 1999, \iaucirc\, 7122

\bibitem[Pignata et al.(2004)]{pignata04} Pignata, G., et al.\ 
2004, \mnras, 355, 178

\bibitem[Puckett \& Sehgal(2000)]{puckett00} Puckett, T., \& Sehgal,
A.\ 2000, \iaucirc\, 7421

\bibitem[Qiao et al.(1997)]{qiao97} Qiao, Q.~Y., Qiu, Y.~L., Li,
W.~D., Hu, J.~Y., Esamdin, A., Wei, J.~Y., Cao, L. \& Gu, Q.~S. 1997,
\iaucirc\, 6775

\bibitem[Qiao et al.(1999)]{qiao99} Qiao, Q.~Y., Wei, J.~Y., Qiu,
Y.~L., \& Hu, J.~Y. 1999, \iaucirc\, 7109

\bibitem[Qiu et al.(1997)]{qiu97} Qiu, Y.~L., Qiao, Q.~Y., Li, W.~D.,
Hu, J.~Y., Esamdin, A., \& Huang, K.~L. 1997, \iaucirc\, 6766

\bibitem[Qiu et al.(1998)]{qiu98} Qiu, Y.~L., Qiao, Q.~Y., \& Hu,
J.~Y.\ 1998, \iaucirc\, 7022

\bibitem[Riess et al.(1996)]{riess96} Riess, A.~G., Press, 
W.~H., \& Kirshner, R.~P.\ 1996, \apj, 473, 88

\bibitem[Riess et al.(1998)]{riess98} Riess, A.~G., et al.\ 
1998, \aj, 116, 1009 

\bibitem[Riess et al.(1999)]{riess99} Riess, A.~G., et al.\ 
1999, \aj, 117, 707

\bibitem[Riess et al.(2001)]{riess01} Riess, A.~G., et al.\ 
2001, \apj, 560, 49

\bibitem[Riess et al.(2004)]{riess04} Riess, A.~G., et al.\ 
2004, \apj, 607, 665

\bibitem[Riess et al.(2005)]{riess05} Riess, A.~G., et al.\ 
2005, \apj, 627, 579

\bibitem[Riess et al.(2007)]{riess07} Riess, A.~G., et al.\ 
2007, \apj, 659, 98

\bibitem[Rudy et al.(2002)]{rudy02} Rudy, R.~J., Lynch, D.~K., Mazuk,
S., Venturini, C.~C., Puetter, R.~C., H{\"o}flich, P.\ 2002, \apj,
565, 413

\bibitem[Salvo et al.(1998)]{salvo98} Salvo, M., et al. 1998,
\iaucirc\, 7037

\bibitem[Schlegel, Finkbeiner, \& Davis(1998)]{schlegel98} Schlegel,
D.,~J., Finkbeiner, D.~P., \& Davis, M. 1998, \apj, 500, 525

\bibitem[Schmidt et al.(1989)]{schmidt89} Schmidt, G., Weymann, R., \&
Foltz, C. 1989, \pasp, 101, 713

\bibitem[Schwartz(1999)]{schwartz99} Schwartz, M.\ 1999, \iaucirc\,
7105

\bibitem[Sollerman et al.(2004)]{sollerman04} Sollerman, J., et al.\
2004, \aap, 428, 555

\bibitem[Stone(1977)]{stone77} Stone, R.~P.~S. 1977, \apj, 218, 767

\bibitem[Stritzinger et al.(2006)]{stritzinger06} Stritzinger, M.,
  Leibundgut, B., Walch, S., \& Contardo, G. 2006, \aap, 450, 241

\bibitem[Stritzinger et al.(2002)]{stritzinger02} Stritzinger, M., et 
al.\ 2002, \aj, 124, 2100

\bibitem[Suntzeff et al.(1999)]{suntzeff99} Suntzeff, N.~B., et al.\
1999, \aj, 117, 1175

\bibitem[Thomas et al.(2004)]{thomas04} Thomas, R.~C., Branch, D.,
Baron, E., Nomoto, K., Li, W., \& Filippenko, A.~V.\ 2004, \apj, 601,
1019

\bibitem[Tonry et al.(2003)]{tonry03} Tonry, J.~L., et al.\ 
2003, \apj, 594, 1


\bibitem[Toth \& Szab{\' o}(2000)]{toth00} Toth, I., \& Szab{\' o},
R.\ 2000, \aap, 361, 63

\bibitem[Villi(1998)]{villi98} Villi, M. 1998, \iaucirc\, 6899

\bibitem[Vink{\' o} et al.(2001)]{vinko01} Vink{\' o}, J., Kiss,
L.~L., Cs{\' a}k, B., F{\H u}r{\' e}sz, G., Szab{\' o}, R., Thomson,
J.~R., \& Mochnacki, S.~W.\ 2001, \aj, 121, 3127

\bibitem[Vink{\' o} et al.(2003)]{vinko03} Vink{\' o}, J., et 
al.\ 2003, \aap, 397, 115

\bibitem[Wade \& Horne(1988)]{wade88}Wade, R.~A., \& Horne,
K.~D. 1988, \apj, 324, 411

\bibitem[Wei \& Li(1998)]{wei98} Wei, J.-Y., \& Li, W.~D. 1998,
\iaucirc\, 6858

\bibitem[Wood-Vasey et al.(2007)]{woodvasey07} Wood-Vasey, W.~M., et
  al. 2007, \apj, 666, 694

\bibitem[Woosley \& Weaver(1986)]{woosley86} Woosley, S., \& Weaver,
  T.~A. 1986, \araa, 24, 205

\bibitem[Yu et al.(2000)]{yu00} Yu, C., Modjaz, M., \& Li, W.~D.\
2000, \iaucirc, 7458

\end{thebibliography}
\end{document}